\documentclass[a4paper,twocolumn,10pt,accepted=2020-08-03]{quantumarticle}
\pdfoutput=1
\usepackage[utf8]{inputenc}
\usepackage[english]{babel}
\usepackage[T1]{fontenc}

\usepackage{graphicx}
\usepackage{xcolor}
\usepackage{amsthm}
\usepackage{amssymb}
\usepackage{amsmath}
\usepackage{physics}

\usepackage[normalem]{ulem}
\usepackage{float}
\usepackage[vlined, ruled]{algorithm2e}  % boxed
\usepackage[export]{adjustbox}  % aligning graphics
\usepackage{subfigure}
\usepackage{soul}

\usepackage[numbers, sort&compress]{natbib}
\newcommand{\mcal}[1]{\mathcal{#1}}		%

\newcommand{\bigO}{\mathcal{O}}

\allowdisplaybreaks % allow equations to be distributed on several pages

\begin{document}
\title{Efficient variational contraction of two-dimensional tensor networks  with a non-trivial unit cell}

\author{A.\ Nietner}
\affiliation{Dahlem Center for Complex Quantum Systems, Freie Universit\"at Berlin, D-14195 Berlin, Germany}
\affiliation{Helmholtz-Zentrum Berlin f\"ur Materialien und Energie, 14109 Berlin, Germany}
%\affiliation{ObstOffice}
\email{anietner@physik.fu-berlin.de}

\author{B.\ Vanhecke}
\affiliation{Department of Physics and Astronomy, University of Ghent, Krijgslaan 281, 9000 Gent, Belgium}
\email{bavhecke.vanhecke@ugent.be}

\author{F.\ Verstraete}
\affiliation{Department of Physics and Astronomy, University of Ghent, Krijgslaan 281, 9000 Gent, Belgium}

\author{J.\ Eisert}
\affiliation{Dahlem Center for Complex Quantum Systems, Freie Universit\"at Berlin, D-14195 Berlin, Germany}
\affiliation{Helmholtz-Zentrum Berlin f\"ur Materialien und Energie, 14109 Berlin, Germany}

\author{L.\ Vanderstraeten}
\affiliation{Department of Physics and Astronomy, University of Ghent, Krijgslaan 281, 9000 Gent, Belgium}

\date{2020-04-29}

\begin{abstract}
Tensor network states provide an efficient class of states that faithfully capture strongly correlated quantum models and systems in classical statistical mechanics. While tensor networks can now be seen as becoming standard tools in the description of such complex many-body systems, close to optimal variational principles based on such states are less obvious to come by. In this work, we generalize a recently proposed variational uniform matrix product state algorithm for capturing one-dimensional quantum lattices in the thermodynamic limit, to the study of regular two-dimensional tensor networks with a non-trivial unit cell. A key property of the algorithm is a computational effort that scales linearly rather than exponentially in the size of the unit cell. We demonstrate the performance of our approach on the computation of the classical partition functions of the antiferromagnetic Ising model and interacting dimers on the square lattice, as well as of a quantum doped resonating valence bond state.
\end{abstract}

\maketitle

\section{Introduction}

Tensor network methods are increasingly becoming a standard tool for studying the physics of strongly-correlated systems, both from the perspective of a theoretical and mathematical understanding of many-body effects as well as for providing a versatile toolbox for numerical simulations\cite{White92,VerstraeteBig,Orus-AnnPhys-2014,SchuchReview,entangAndTNS,Handwaving}. 
In the context of one-dimensional quantum physics, \emph{matrix product states (MPS)} 
have been identified to parametrize the low-energy states of gapped local Hamiltonians, 
in fact provably so. MPS-based algorithms allow for efficient simulations of static and dynamic properties of spin chains in various facets \cite{Schollwoeck2011} and allow for precise numerical analysis of symmetry protected topological order in one dimension \cite{Pollmann2012, Schuch2011,nietner2017}. Two-dimensional quantum systems can be simulated using 
instances of \emph{projected entangled-pair states (PEPS)}. Indeed, the PEPS toolbox is increasingly capturing ground state \cite{Corboz2014a, Corboz2014b,Vanderstraeten2016, kshetrimayum2019}, dynamic \cite{Czarnik2019a, Hubig2019,MBL2D}, spectral \cite{Vanderstraeten2015, Vanderstraeten2019b}, and finite-temperature \cite{Kshetrimayum2019thermal, Czarnik2012}, properties as well as the simulation of open system dynamics \cite{kshetrimayum2016steady} of quantum spins or electrons in two dimensions. In the field of classical statistical mechanics, tensor networks provide a natural way of simulating e.g. critical and/or frustrated systems in two \cite{Nishino1995} and three \cite{Nishino2001, Vanderstraeten2018} dimensions, which are notoriously difficult for standard sampling methods.
\par When applying tensor networks to two- and three-dimensional systems, both quantum and classical, the computational bottleneck always consists of the contraction of a two-dimensional tensor network. In the easiest case, one considers a translation-invariant tensor network on a regular lattice in the thermodynamic limit, i.e. the network consists of a single tensor that is repeated over an infinite lattice. In the past, different algorithms have been proposed for this fundamental task, which can be subdivided into three main approaches: (i) Real-space renormalization-group methods \cite{Levin2007, Xie2012, Evenbly2015, Yang2017, Bal2017, Hauru2018} rely on coarse-graining transformations on the level of the tensors that make up the network, such that global properties of the tensor network can be efficiently computed. (ii) In corner transfer matrix methods \cite{Baxter1968, Baxter1978, Nishino1996, Nishino1997, Orus2009, Corboz2010}, part of the network is represented by effective corner tensors which are obtained by an iterative growing of the environment and truncation of the tensors. (iii) Boundary methods aim to approximate the fixed point of the row-to-row transfer matrix as an MPS, such that this MPS represents an effective representation of half of the network; different algorithms can be used to find the fixed point such as the density-matrix renormalization group \cite{White92, Nishino1995}, the time-evolving block decimation \cite{Vidal03,Orus2008}, the tensor ring decomposition \cite{ShiJu2016, ran2018review} or the VUMPS algorithm \cite{ZaunerStauber2018, Haegeman2016, Vanderstraeten2019a}.

The two former approaches rely on power methods to find a fixed point for the contraction of an infinite tensor network, whereas the latter two iterate local self-consistency relations. In particular variational
MPS-tangent-space methods such as VUMPS
can exploit more advanced solvers for the leading eigenvector of the transfer matrix. This property leads to a significant speed-up for the VUMPS algorithm as compared to power methods when critical or close-to-critical tensor networks are considered \cite{Fishman2018, ZaunerStauber2018}.

In many applications, the relevant two-dimensional tensor network cannot be chosen to be translation invariant, but rather consists of a larger unit cell of different tensors that are repeated over the infinite lattice. In other scenarios, the tensor network itself is translation-invariant but the lattice symmetry is spontaneously broken. In both cases, an algorithm with uniform tensors can not be used for the contraction. Whereas corner transfer matrix approaches have been extended to the case of larger unit cells \cite{Corboz2011,Corboz2014b}, the variational boundary-MPS methods have not been formulated in this more general setting. In this work, we show that this generalization of the VUMPS algorithm is, in fact, a very natural one and leads to an algorithm with a complexity that scales linearly with the size of the non-trivial unit cell.

\section{Set-up}

In this  section we review the basic ingredients that allow for the contraction of translation-invariant 
two-dimensional tensor networks using the VUMPS algorithm \cite{ZaunerStauber2018}, setting the stage for the multi-site version 
in the next section.

\subsection{Two-dimensional tensor networks}

Two-dimensional tensor networks are most naturally obtained in the context of two-dimensional statistical mechanics, where they appear as a representation of the partition function of lattice spin models with local interactions. Indeed, suppose we have a system of spins $s_i=\pm1$ on a regular, say square, lattice with nearest-neighbour interactions 
\begin{equation}\label{hamiltonian}
\mathcal{H} = \sum_{\langle i,j \rangle} H(s_i,s_j) .
\end{equation}
The partition function for this model is given by
\begin{align}
\mathcal{Z} &= \lim_N\sum_{s\in\{\pm1\}^N} \mathrm{e}^{-\beta\mathcal{H}(s)} \nonumber\\
&= \lim_N\sum_{s\in\{\pm1\}^N} \prod_{\langle i,j \rangle} \mathrm{e}^{-\beta H(s_i, s_j) }.
\end{align}
We can now write this partition function as a tensor network by placing a local Boltzmann weight represented by the symmetric matrix $t$
\begin{equation}
\label{af_ising_tensors2}
\begin{gathered}
\includegraphics[valign=c]{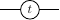}			
= \mathrm{e}^{-\beta H(s_1,s_2)}
\end{gathered}
\end{equation}
on each link on the lattice, placing a $\delta$ tensor
\begin{equation}
\begin{gathered}
\includegraphics[valign=c]{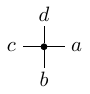}	
= \left\{
\mqty{\,1, \qquad & a=b=c=d \\ 
\,0, \qquad & \text{else}} \right.,
\end{gathered}
\end{equation}
on each site, and contracting all connected indices. In this way, we arrive at
\begin{equation}\label{tensor_network_expl}
\begin{gathered}
\mcal{Z} =
\includegraphics[valign=c]{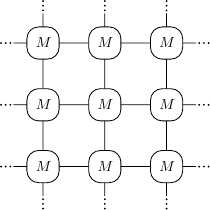}			
\end{gathered}
\end{equation}
with
\begin{equation}
\label{af_ising_tensors}
\begin{gathered}
\includegraphics[valign=c]{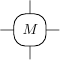}			
= 
\includegraphics[valign=c]{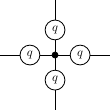}					
\end{gathered}
\end{equation}
where $q^2=t$. A local expectation value of an observable $O$ with values $O_j$, $j=\pm1$, at site $i$ is given by changing one tensor in this tensor network, i.e.,
\begin{equation}\label{expectation_value_expl}
\begin{gathered}
\langle O^i\rangle =
\frac{1}{\mathcal{Z}} 
\includegraphics[valign=c]{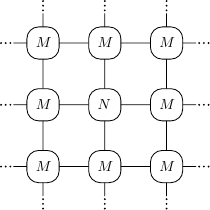}		
\end{gathered}
\end{equation}
where the new tensor $N$ is given by
\begin{equation}
\label{af_ising_tensors3}
\begin{gathered}
\includegraphics[valign=c]{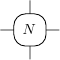}			
= 
\includegraphics[valign=c]{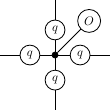}					
\end{gathered}
\end{equation}
and is placed at site $i$. Similarly one can represent generic $n$-point functions by placing the $n$ defect tensors in the form of Eq.~\eqref{af_ising_tensors3} at the corresponding sites in the partition function Eq.~\eqref{tensor_network_expl}.

Two-dimensional tensor networks also show up as the norm or local expectation values of two-dimensional PEPS. Then, the elementary four leg tensor $M$ is given as the sandwich of the PEPS tensor and its conjugate 
\begin{equation}
\label{peps_mpo}
\begin{gathered}
\includegraphics[valign=c]{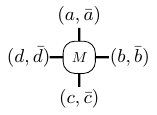}			
= 
\includegraphics[valign=c]{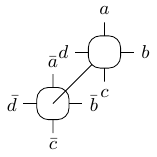}					
\end{gathered}
\end{equation}
where the thick leg on the left hand side corresponds to the tensor product of the two thin legs of the respective side of the PEPS tensor. 

To contract two dimensional tensor networks it is natural to use the fact that topological two dimensional systems (trivial or non-trivial) are in general believed to admit a gapped boundary \cite{bauer2019, hastings2006}. This is, there exists a many body state vector $\ket{\psi}$ with exponentially decaying correlations such that
\begin{equation}
\label{def_eq_boundary}
\begin{gathered}
\includegraphics[valign=c]{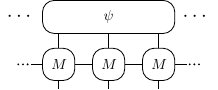}	
=\lambda
\includegraphics[valign=c]{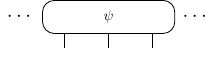}			  
\end{gathered}.
\end{equation}
Note that similar to Ref.~\cite{bauer2019}, we use the notion of a boundary interchangeably for the boundary of the lattice as well as for the boundary state living on the boundary. If one is able to efficiently compute and manipulate this boundary one can use the fixed point equation~\eqref{def_eq_boundary} in order to reduce the contraction of the two-dimensional system to the contraction of a one-dimensional system. 

\subsection{Matrix product states in the thermodynamic limit}\label{MPS}

\emph{Matrix product states (MPS)} are a class of states that can be efficiently contracted and that provably capture the local properties of gapped boundaries arbitrarily well \cite{Schollwoeck2011,Brandao2015, schuch2017, dalzell2019, huang2019}. Therefore, MPS are a powerful tool for approximating boundaries of two-dimensional tensor networks in the thermodynamic limit in an efficient way. An MPS is fully determined by a three-leg tensor $A=(A_{\alpha, i, \beta})$ with dimensions $(\chi, d,\chi)$ where $\chi$ is called the bond dimension and $d$ the physical dimension of the MPS. The boundary represented by this tensor can be written in the thermodynamic limit as
\begin{equation}
\begin{gathered}
\ket{\psi(A)}
=
\includegraphics[valign=c]{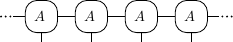}		
\end{gathered}\;.
\end{equation}
The representation of a state by a three-leg tensor is not unique due to the gauge degrees of freedom of introducing an identity in the form of a pair of matrices $X\cdot X^{-1}$ on a bond. This gauge freedom can be exploited to choose a canonical form for the tensors. In the so-called mixed canonical gauge one fixes a reference site $i$ and all tensors to the left of $i$ are represented in the left canonical gauge $A_L$, and all tensors to the right of $i$ are represented by the right canonical gauge $A_R$, i.e.
\begin{equation}
\label{mixed_canonical_form}
\begin{gathered}
\ket{\psi(A_L, A_C, A_R)}
=	
\includegraphics[valign=c]{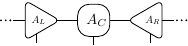}			
\end{gathered}
\end{equation}
where we have introduced a center site tensor $A_c$, represented as 
\begin{equation}
\label{triv_center_definition}
\begin{gathered}
\includegraphics[valign=c]{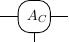}
=
\includegraphics[valign=c]{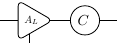}
=
\includegraphics[valign=c]{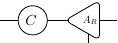}		
\end{gathered},
\end{equation}
with $C=(C_{\alpha ,\beta})$ the matrix that gauge transforms $A_L$ to $A_R$ and whose singular values are the Schmidt values of the state corresponding to the cut through that bond. The left and right canonical tensors are defined as the tensors representing the same state as $A$ with the additional properties of
\begin{equation}
\label{left_gauge}
\begin{gathered}
\includegraphics[valign=c]{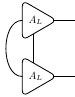}
=
\includegraphics[valign=c]{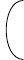}		
\end{gathered}
\end{equation}
and
\begin{equation}
\label{right_gauge}
\begin{gathered}
\includegraphics[valign=c]{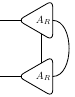}
=
\includegraphics[valign=c]{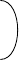}		
\end{gathered}\, ,
\end{equation}
respectively. In other words, the tensors $A_L$ and $A_R$ are isometries\footnote{Throughout this paper we have the convention that MPS tensors with their physical leg pointing upwards are the complex conjugated version of those with their legs pointing downwards.}.

The set of MPS tensors for a fixed bond dimension $\chi$ modulo gauge equivalence defines the manifold of MPS $\mathcal{M}_\chi$. Through the map $\ket{\psi(\cdot)}$ that embeds the tensors into the many body Hilbert space, one obtains a metric on $\mathcal{M}_\chi$ and a notion of a tangent space projector that projects states from the Hilbert space onto the tangent space of the manifold \cite{Haegeman2013, Vanderstraeten2019a}, cf.~\ref{vumps_sec}.

\subsection{Approximating center tensors}

Motivated by the VUMPS algorithm, it is convenient to have a means of finding a MPS approximation to a given pair of tensors $A_C$ and $C$. In principle, a MPS is fully defined via $A_L$ or $A_R$ alone. However, one can define a MPS via $A_C$ and $C$, where $A_L$ and $A_R$ can  be found by using Eq.~\eqref{triv_center_definition} and inverting the matrix $C$. Note, however, that not every pair consisting of a three- and a two-leg tensor exactly represents a MPS, as solving Eq.~\eqref{triv_center_definition} via inverting $C$ not necessarily gives rise to a left or right isometric tensor. Given two tensors $A_C$ and $C$ corresponding to a MPS, one can find the isometric tensors $A_L$ and $A_R$ that realize Eq.~\eqref{triv_center_definition} by a singular-value decomposition (or, likewise, by a polar decomposition) \cite{ZaunerStauber2018} 
\begin{equation}
\begin{gathered}
\includegraphics[valign=c]{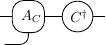}	
=	
\includegraphics[valign=c]{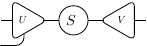}	\end{gathered}
\end{equation}
\begin{equation}
\label{truncation_a_l}
\begin{gathered}
\includegraphics[valign=c]{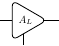}	
=	
\includegraphics[valign=c]{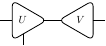}	
\end{gathered}\;,
\end{equation}
and similarly for $A_R$
\begin{equation}
\begin{gathered}
\includegraphics[valign=c]{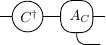}	
=	
\includegraphics[valign=c]{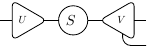}	\end{gathered}\;
\end{equation}
and
\begin{equation}
\begin{gathered}
\label{truncation_a_r}
\includegraphics[valign=c]{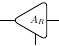}	
=	
\includegraphics[valign=c]{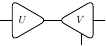}	\end{gathered}\,. 
\end{equation}
Obtaining isometries $A_L$ and $A_R$ from some pair $A_C$ and $C$ (not necessarily explicitly representing a MPS) gives rise to gauging errors $\epsilon_{L}$ and $\epsilon_{R}$ as
\begin{equation}
\begin{gathered}
\epsilon_L
=	
\left|\left|
\includegraphics[valign=c]{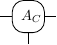}	
-
\includegraphics[valign=c]{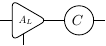}	
\right|\right|_2
\end{gathered}
\end{equation}
and
\begin{equation}
\begin{gathered}
\epsilon_R
=	
\left|\left|
\includegraphics[valign=c]{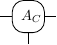}	
-
\includegraphics[valign=c]{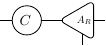}		
\right|\right|_2
\end{gathered}.
\end{equation}

\subsection{Variational optimization for uniform matrix product states (VUMPS)}\label{vumps_sec}

The VUMPS algorithm \cite{ZaunerStauber2018} is a fixed point iteration method for finding the boundary of a two-dimensional tensor network. The desired fixed point equation is obtained starting from expressing the eigenvalue equation for the boundary state Eq.~\eqref{def_eq_boundary} in terms of MPS as
\begin{equation}
\label{fixed_point_mpo_equation}
\begin{gathered}
\includegraphics[valign=c]{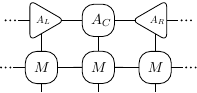}	
\approx	\lambda
\includegraphics[valign=c]{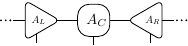}			
\end{gathered}.
\end{equation}
The approximation sign in Eq.~\eqref{fixed_point_mpo_equation} is due to the MPS approximation of the boundary, and signifies that we aim at finding an MPS that approximates this equation in an optimal way.
% ausformulieren
Here, we chose tho following optimality condition: We interpret the set of MPS with a given bond dimension $\chi$ as a variational manifold $\mathcal{M}_\chi$. Next, we observe that the MPS on the left hand side of Eq.~\eqref{fixed_point_mpo_equation} has bond dimension $\chi D$, with $D$ the bond dimension of the MPO, whereas the MPS on the right hand side has bond dimension $\chi$. Then, optimality with respect to the variational manifold $\mathcal{M}_\chi$ implies that the left hand side variationally truncated to $\chi$ should equal the right hand side.
% ausformulieren ende
This optimality condition can be reformulated as saying that we look for the MPS for which the error made in Eq.~\eqref{fixed_point_mpo_equation} is orthogonal (in Hilbert space) to the tangent space on the manifold. Put differently, the tangent space projector of the manifold applied to the equation should vanish to guarantee the optimal solution within the variational manifold.

The tangent space projector $\mathcal{P}_A$ at $A$ (projecting any state in the many body Hilbert space onto its overlap with the tangent space at the state parametrized by $A$) can be graphically written as \cite{Haegeman2016}
\begin{equation}
\begin{gathered}
\label{tangent_space_projector}
\mathcal{P}_{A} = \\\sum_{i\in\mathbb{Z}}
\includegraphics[valign=c]{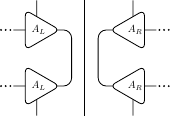}	
-
\includegraphics[valign=c]{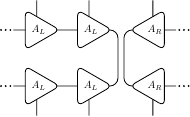}	
\\=\sum_{i\in\mathbb{Z}}
\includegraphics[valign=c]{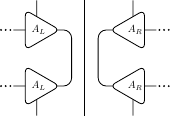}	
-
\includegraphics[valign=c]{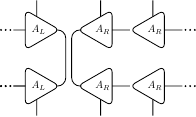}	
\end{gathered},
\end{equation}
where $i$ denotes the site in the lattice at which the picture is centred. Introducing the tensors
\begin{equation}
\begin{gathered}
\frac{1}{\mathcal{N}}
\includegraphics[valign=c]{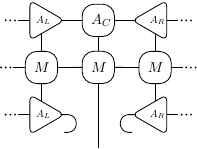}	
=	 
\includegraphics[valign=c]{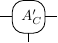}			
\end{gathered}
\end{equation}
and
\begin{equation}
\begin{gathered}
\frac{1}{\mathcal{N}'}
\includegraphics[valign=c]{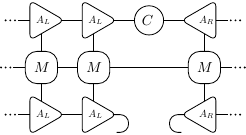}	
=	
\includegraphics[valign=c]{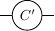}			
\end{gathered}
\end{equation}
where the diverging $\mathcal{N}$ and $\mathcal{N}'$ denote the normalizations counteracting the non-normalized tensors $M$, one finds \cite{ZaunerStauber2018} that the vanishing of the tangent space projector on Eq.~\eqref{fixed_point_mpo_equation} is equivalent to Eq.~\eqref{triv_center_definition} together with
\begin{equation}
\begin{gathered}
\includegraphics[valign=c]{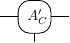}
=
\includegraphics[valign=c]{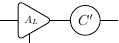}
=
\includegraphics[valign=c]{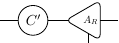}		
\end{gathered}\;.
\end{equation}
Moreover, because the gauge transformation $C$ transforming $A_L$ and $A_R$ into each other is unique up to a factor, it must hold that $A_C'\propto A_C$. This leads to the VUMPS equations
\begin{equation}
\begin{gathered}
\label{vumps_1}
\frac{1}{\mathcal{N}}
\includegraphics[valign=c]{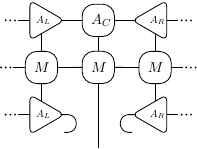}	
=	 \tau_{A_C}
\includegraphics[valign=c]{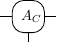}			
\end{gathered}
\end{equation}
and
\begin{equation}
\label{vumps_2}
\begin{gathered}
\frac{1}{\mathcal{N}'}
\includegraphics[valign=c]{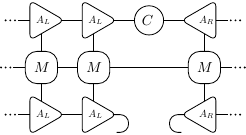}	
=	\tau_C
\includegraphics[valign=c]{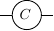}			
\end{gathered}\;.
\end{equation}
Together with Eq.~\eqref{triv_center_definition}, Eq.~\eqref{vumps_1} and \eqref{vumps_2} fully characterize the fixed point. 
To solve this equation we run into one obvious problem: The isometries $A_L$ and $A_R$ are not known unless we 
have identified the full MPS, which we do not know if we want to solve for $\ket{\psi(A)}$. Still, we can use the framework from above to iteratively approximate the desired fixed point.

Starting from a random MPS and solving this equation once for $A_C$ and $C$ respectively we can use Eq.~\eqref{left_gauge} and \eqref{right_gauge} to find a new MPS approximating this pair of tensors with a gauging error $\max\{\epsilon_L, \epsilon_R\}$. This update implicitly defines a non-linear map in the MPS manifold. Iterating the procedure we will eventually converge to the fixed point of this map represented by a pair of $A_C$ and $C$ with vanishing gauging error. This fixed point will then correspond to the optimal approximation to the boundary of the tensor network within the MPS manifold. The gauging errors $\epsilon_L$ and $\epsilon_{R}$ correspond to the norm of the gradient of $\ev{M}{\psi(A)}$ with respect to $A$. Similar as in DMRG and other contraction methods it is possible to get stuck in a local minimum, which in practice however is rarely the case.

\section{Contracting tensor networks with non trivial unit cell}

In this section we will show how the above mentioned method can be extended to the case of a non-trivial unit cell with the computational effort of the algorithm growing only linearly rather than exponentially in the unit cell size.

\subsection{MPS with non-trivial unit cell}\label{nt_mps}

In order to deal with tensor networks with a non-trivial unit cell it is convenient to first introduce MPS with a non-trivial unit cell. A MPS with a non-trivial unit cell of size $n$ is given by the data of the $n$ tensors $A_i$, $i=1,\dots,n$, 
of shape $(\chi, d, \chi)$, or equivalently by a single four leg tensor $A$ of shape $(\chi, d, \chi, n)$, where the $A_i$ are repeated cyclically in order to construct the MPS
\begin{equation}\label{ntmps}
\begin{gathered}
\ket{\psi(A)}
=
\includegraphics[valign=c]{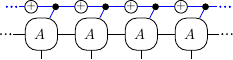}		
\end{gathered} 
\end{equation}
Here, the blue index corresponds to $i=1,\dots,n$ and the $+$ matrix is the raising operator mapping the $i$-th basis vector to the $i+1$-th basis vector modulo $n$.
Note that throughout this work the raising operator will be used such that it is either acting from the left to the right or acting from top to bottom. The dimension will be clear from the context. Moreover, if it is clear from the context that we are dealing with a non-trivial unit cell MPS we might drop the dependency on $i$ (the blue leg) for the sake of a clearer picture. Note that the state as in Eq.~\eqref{ntmps} is a superposition of all unit cell shifts of the MPS with a non trivial unit cell. However, fixing the $x$-channel to a specific value at any point (eg via an additional leg at one of the delta tensors) one can fix to a specific MPS with a non-trivial unit cell.

Similar as for standard MPS  we can bring $\ket{\psi(A)}$ to a canonical form defined with tensors $A_C$ and $C$ (now both having an additional leg encoding the position in the unit cell)
\begin{equation}
\label{center_definition}
\begin{gathered}
\includegraphics[valign=c]{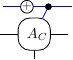}
=
\includegraphics[valign=c]{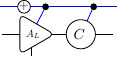}
=
\includegraphics[valign=c]{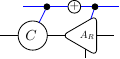}		
\end{gathered}
\end{equation}
where again the left and right canonical tensors $A_L$ and $A_R$ fulfill (\ref{left_gauge}) and (\ref{right_gauge}) at 
each $i=1,\dots,n$, separately. A multi-site MPS can, therefore, be represented as
\begin{equation}
\label{nt_mixed_canonical_form}
\begin{gathered}
\ket{\psi(A_L, A_C, A_R)}
=	
\includegraphics[valign=c]{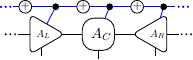}			
\end{gathered}\;.
\end{equation}

The formulas for finding $A_L$ and $A_R$ from a given $A_C$ and $C$ similar to Eq.~\eqref{truncation_a_l} and \eqref{truncation_a_r} for MPS with a non-trivial unit cell will turn out to be very convenient in order to generalize the VUMPS algorithm. It is straightforward to see that they can be generalized to
\begin{equation}
\begin{gathered}
\includegraphics[valign=c]{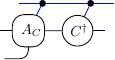}	
=	
\includegraphics[valign=c]{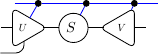}	\end{gathered}
\end{equation}
\begin{equation}
\label{nt_truncation_a_l}
\begin{gathered}
\includegraphics[valign=c]{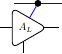}	
=	
\includegraphics[valign=c]{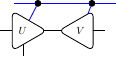}	
\end{gathered}
\end{equation}
and
\begin{equation}
\begin{gathered}
\includegraphics[valign=c]{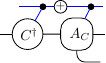}	
=	
\includegraphics[valign=c]{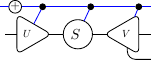}	\end{gathered}
\end{equation}
\begin{equation}
\begin{gathered}
\label{nt_truncation_a_r}
\includegraphics[valign=c]{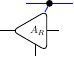}	
=	
\includegraphics[valign=c]{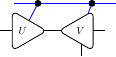}	
\end{gathered}\;.
\end{equation}

\subsection{Non-trivial unit cell VUMPS}\label{section:non_trivial_vumps}

Let us now generalize the VUMPS algorithm to tensor networks with a non-trivial unit cell. As before we assume a tensor network on a square lattice now with a unit cell of shape $(n_x, n_y)$ with $D$ denoting the bond dimension of the tensor network. We can write this as the consecutive application of $n_y$ MPOs $M_{n_y}\cdots M_1$, each with a substructure of a unit cell of length $n_x$. Again, we assume the existence of a gapped boundary of the tensor network that we approximate by an MPS $\ket{\psi(A)}$ with a non-trivial unit cell. Due to the non-trivial unit cell, $\ket{\psi(A)}$ gets mapped onto itself only after the consecutive application of the $n_y$ MPOs. As this argument is independent on the cyclic order of the $n_y$ MPOs there exist $n_y$ boundaries $\ket{\psi(A_1)}$, $\dots$, $\ket{\psi(A_{n_y})}$ for which the fixed point equation reads
\begin{equation}
\label{MPO_eigenvalue_equation}
\begin{gathered}
\includegraphics[valign=c]{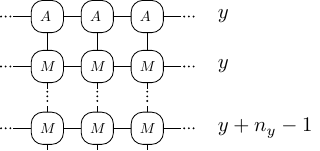}
\\[10pt]
\approx 
\lambda_y
\includegraphics[valign=c]{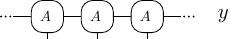}		
\end{gathered}
\end{equation}
where we have dropped the dependencies of the tensors on the $x$ and $y$ coordinate for the sake of simplicity of the picture (as with the substructure of a non-trivial unit cell  MPS we can encode the $y$ dependency of the tensor in an additional leg for the MPS and MPO tensors which, in case of the $y$ coordinate, we will encode by a red coloured leg in the following, cf. Appendix \ref{appendix:non_trivial_tensor_networks}). It is easy to see that $\lambda_y$ is constant in $y$: if $M_{y+n_y}\cdots M_{y}\ket{\psi} = \lambda_y\ket{\psi}$ then it must hold that $M_y M_{y+n_y}\cdots M_{y-1}(M_{y}\ket{\psi}) = \lambda_y(M_{y}\ket{\psi})$. More generally, the full spectrum of $M_{\sigma(y+n_y)}\cdots M_{\sigma(y)}$ is the same for all cyclic permutations $\sigma$.

One way of solving (\ref{MPO_eigenvalue_equation}) is to block the MPOs together and thereby reduce the problem to a tensor network with a trivial unit cell for which one can apply the VUMPS algorithm from the previous section. However, this comes at the cost of increasing the bond dimension of the MPO to $D^{n_y}$ (the blocking along the $x$ direction can be avoided with a linear overhead as shown in Ref.~\cite{ZaunerStauber2018}). Let us therefore introduce a simple trick to solve (\ref{MPO_eigenvalue_equation}) with only a linear instead of an exponential overhead in $n_y$.

The general idea is as follows. Given $n$ square matrices $X_1, \dots, X_n$ of dimension $k\times k$ such that the product $X_{y+n}\cdots X_y$ is diagonalizable. We are looking for the $n$ vectors $v_y$ such that
\begin{equation}
\label{blocked_eigenvalue_equation}
\begin{gathered}
\includegraphics[valign=c]{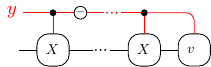}
=
\lambda
\includegraphics[valign=c]{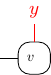}		
\end{gathered}
\end{equation}
where $-$ is the inverse of the $+$ operator, $\lambda$ is the eigenvalue corresponding to the consecutive application of all the $X_y$, the black dots are $\delta$ tensors and the dots indicate that we multiply over all the $n_y$ matrices. Eq.~\eqref{blocked_eigenvalue_equation} can be solved either by blocking the $n$ matrices to a single matrix, or we use the fact that it holds that
\begin{equation}
\label{pre_eigenvector_equation}
\begin{gathered}
\includegraphics[valign=c]{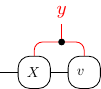}
= 
\includegraphics[valign=c]{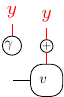}		
\end{gathered}
\end{equation}
where we have written $\gamma_y$ as a vector in $y$ where $\gamma_{y+n}\cdots \gamma_y=\lambda$ with lambda is the respective eigenvalue of the blocked matrices. It is straightforward to check (\ref{pre_eigenvector_equation}) by simply substituting it into (\ref{blocked_eigenvalue_equation}). We can rewrite (\ref{pre_eigenvector_equation}) as
\begin{equation}\label{tm_gap_definition}
\begin{gathered}
\includegraphics[valign=c]{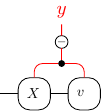}
= 
\includegraphics[valign=c]{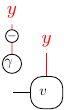}		
\end{gathered}
\end{equation}
Finally, absorbing the $\gamma_y$ into the norm of the $v_{y+1}$ we obtain an eigenvector equation for the vector $v$ in the vector space corresponding to the tensor product space of the two legs of $v$
as
\begin{equation}
\label{pvumps_core}
\begin{gathered}
\includegraphics[valign=c]{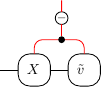}
= \tilde{\gamma}
\includegraphics[valign=c]{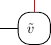}		
\end{gathered}
\end{equation}
with
% \begin{equation}
% \begin{gathered}
% \includegraphics[valign=c]{tikz_73_73.pdf}		
% \propto
% \includegraphics[valign=c]{tikz_74_74.pdf}		.
% \end{gathered}
% \end{equation}
\begin{equation}\label{eq:def-tilde-v}
\tilde{v}_y=\frac{\left(\prod_{i=0}^{y-1}\gamma_i\right)^{(n-y+1)/n}}{\left(\prod_{i= y}^n\gamma_i\right)^{(y-1)/n}}\cdot v_{y}\,,
\end{equation}
where we set $\gamma_0=1$ and with $\gamma_y$, $y\geq1$, as before.

Eq.~\eqref{pvumps_core} is an eigenvector equation solving for all $v_i$ at the same time. Assuming an iterative solver this corresponds to increasing the complexity from $\bigO(k)$ to $\bigO(nk)$, which in turn corresponds to a linear overhead in the scaling parameter $n$. While this is not necessary for matrices, as blocking increases the complexity only linearly, too, it is crucial for MPOs due to the non-trivial bond dimension. Finally, the crucial observation is that we can re-obtain the sought-after $v_y$ from the solution $\tilde{v}$ simply by normalizing the respective $\tilde{v}_y$ (c.f. Eq.~\eqref{eq:def-tilde-v}).

Let us now do the same on the level of MPOs. We rewrite Eq.~\eqref{MPO_eigenvalue_equation} similar to Eq.~\eqref{pre_eigenvector_equation} as
\begin{equation}
\label{pre_MPO_eigenvalue_equation}
\begin{gathered}
\includegraphics[valign=c]{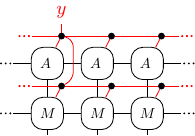}
\approx 
\gamma_y
\includegraphics[valign=c]{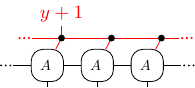}		
\end{gathered}
\end{equation}
where we made the $y$ dependence on the tensors explicit for clarity. This is the analogue of (\ref{pre_eigenvector_equation}) on the level of tensor networks. Hence, we can use the same arguments as before to obtain the tensor network 
analogue of (\ref{pvumps_core})
\begin{equation}
\label{local_MPO_eigenvalue_equation}
\begin{gathered}
\includegraphics[valign=c]{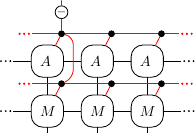}
\approx 
\tilde{\gamma}
\includegraphics[valign=c]{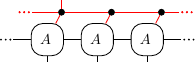}		
\end{gathered}
\end{equation}
Similar as before, we project down this equation in order to obtain a fixed-point equation on the level of the tensors just as in \ref{vumps_sec}. 

We start from Eq.~\eqref{local_MPO_eigenvalue_equation} and enforce the tangent space projectors corresponding to the states of each $y$ on the right hand side to vanish. This corresponds to the application of the generalized tangent space projector as in Appendix \ref{appendix:non_trivial_geometry}. Combining this with the gauge constraints Eq.~\eqref{center_definition} and the fact that the $n_y$ MPS with a fixed $x$- and $y$-index are injective, we obtain the equations for $A_C^{y+1}$ and $C^{y+1}$ respectively
\begin{equation}
\label{local_AC_eigenvalue_equation}
\begin{gathered}
\frac{1}{\mathcal{N}}
\includegraphics[valign=c]{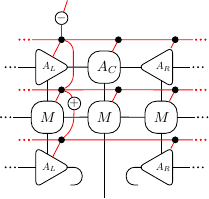}
= 
\tau_{A_C}
\includegraphics[valign=c]{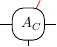}		
\end{gathered}
\end{equation}
and
\begin{equation}
\label{local_C_eigenvalue_equation}
\begin{gathered}
\frac{1}{\mathcal{N}'}
\includegraphics[valign=c]{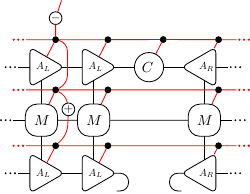}	
=	
\tau_{C}
\includegraphics[valign=c]{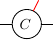}			
\end{gathered}\;.
\end{equation}
A more detailed derivation of these fixed-point equations can be found in Appendix~\ref{appendix:non_trivial}. The variational optimality can be easily understood as shown in Appendix \ref{variational_argument}.

\begin{algorithm*}
\SetAlgoVlined
\SetInd{1em}{1em} 
\DontPrintSemicolon

\SetKwFunction{ENvs}{environment\_terms}
\SetKwProg{Fn}{Function}{:}{}
\Fn{\ENvs{$(M_{x,y})_{x,y}$, $(A_L^{x,y})_{x,y}$, $(A_R^{x,y})_{x,y}$, $\epsilon$}}{
$(L_{x,y})_{x,y} \longleftarrow$ compute left eigenvector according to (\ref{l_channel_fixed_Point})  \;
$(R_{x,y})_{x,y} \longleftarrow$ compute left eigenvector according to (\ref{r_channel_fixed_Point})  \;
\KwRet $(L_{x,y})_{x,y}$ and $(R_{x,y})_{x,y}$\;
}
\;

\SetKwFunction{PAPAC}{apply\_h\_a\_c}
\Fn{\PAPAC{$(A_C^{x,y})_{x,y}$}}{
$(A_C^{x,y})_{x,y} \longleftarrow $ update according to (\ref{local_h_a_c})\;
\KwRet $(A_C^{x,y})_{x,y}$\;
}

\;
\SetKwFunction{PAPC}{apply\_h\_c}
\Fn{\PAPC{$(C^{x,y})_{x,y}$}}{
$(C^{x,y})_{x,y} \longleftarrow $ update according to (\ref{local_h_c})\;
\KwRet $(C^{y})_y$\;
}

\caption{Explicit terms of the local Hamiltonians in the parallel implementation.}
\label{p_environment_algorithms}
\end{algorithm*}

\subsection{The algorithm}\label{section:algorithm}

We obtain the $n_y$ coupled local equations for the $A_C$ and $C$ tensors Eq.~\eqref{local_AC_eigenvalue_equation} and \eqref{local_C_eigenvalue_equation} which, together with Eq.~\eqref{nt_truncation_a_l} and \eqref{nt_truncation_a_r} implicitly defines a flow through the coupled MPS manifolds of the $n_y$ boundaries. This flow can be integrated  just as in \ref{vumps_sec}. To do so we solve the local equations Eq.~\eqref{local_AC_eigenvalue_equation} and \eqref{local_C_eigenvalue_equation} for an initial set of MPS, update all boundaries according to the just obtained solutions for $A_C$ and $C$ using Eq.~\eqref{nt_truncation_a_l} and \eqref{nt_truncation_a_r}. Again, derive the new local equations according to the current set of boundaries. Iterate this procedure until convergence is reached.

Let us now phrase this in explicit algorithmic form. We can compute the left and right channels in (\ref{local_AC_eigenvalue_equation}) and (\ref{local_C_eigenvalue_equation}) explicitly by using the leading left and right eigenvectors of the transfer matrices respectively. To this end, we can either use (\ref{pvumps_core}) for the consecutive application of the transfer matrices. However, Eq.~\eqref{l_channel_fixed_Point} and Eq.~\eqref{r_channel_fixed_Point} are already optimal in complexity because we are dealing with matrices rather than MPOs. On the other hand, for sufficiently small gaps $\gamma_i$ in the transfer matrices (cf. Eq~\eqref{tm_gap_definition}) Eq~\eqref{pvumps_core} leads to more stability. Moreover, using an iterative solver we can exploit the tensor-network structure to reduce the memory allocation. Thus one obtains a computational effort of $\bigO(D^2\chi^3 + D^4\chi^2)$ compared to $\bigO(D^2\chi^4)$, which can further be reduced by parallelizing\footnote{In terms of the PEPS bond dimension $D_{P}$ those complexities correspond to $\bigO(D_{P}^4\chi^3+D_{P}^8\chi^2)$ compared to $\bigO(D_{P}^4\chi^4)$.}. The equation for the $n_y\times n_x$ left eigenvectors is 
\begin{equation}
\label{l_channel_fixed_Point}
\begin{gathered}
\includegraphics[valign=c]{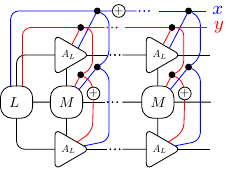}	
=	
\alpha_{x,y}^L\;\;
\includegraphics[valign=c]{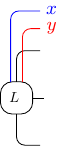}			
\end{gathered}
\end{equation}
where we have drawn the dependence of the tensors on $x$ and $y$ explicitly for the sake of clarity. The dots imply the matrix product over the $n_x$ transfer matrices contained in one unit cell of the $y$-th row. Similarly, the equations for the right eigenvectors are
\begin{equation}
\label{r_channel_fixed_Point}
\begin{gathered}
\includegraphics[valign=c]{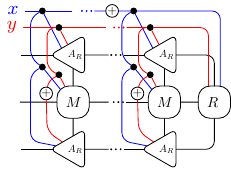}	
=	
\alpha_{x,y}^R\;
\includegraphics[valign=c]{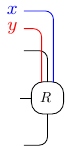}			
\end{gathered}\;.
\end{equation}
Note, while we need to solve this equation for all $y$ we can fix the $x$ value and obtain the remaining environment tensors via applying the respective transfer matrices (followed by a normalization).

We can now use the $(x,y)$-dependent environments $(L_{x,y})_{x,y}$ and $(R_{x,y})_{x,y}$ in order to define the  application of the local fixed point equations $h_{A_C}$ and $h_C$ corresponding to (\ref{local_AC_eigenvalue_equation}) and (\ref{local_C_eigenvalue_equation}) that we will use to iteratively solve for the tensors $A_C$ and $C$ 

\begin{equation}
\label{local_h_a_c}
\begin{gathered}
\includegraphics[valign=c]{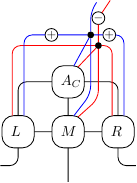}	
=	
\lambda_{A_C}\;
\includegraphics[valign=c]{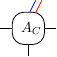}			
\end{gathered}
\end{equation}
and

\begin{equation}
\label{local_h_c}
\begin{gathered}
\includegraphics[valign=c]{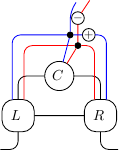}	
=	
\lambda_{C}\;
\includegraphics[valign=c]{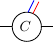}			
\end{gathered}
\end{equation}

Let us recapitulate the algorithm: To start we initialize a set of $n_y$ MPS with a unit cell of size $n_x$, each of these MPS corresponding to a boundary. Then we derive the local equations $h_{A_C}$ (cf.~ (\ref{local_h_a_c})) and $h_c$ (cf.~(\ref{local_h_c})) using (\ref{l_channel_fixed_Point}) and (\ref{r_channel_fixed_Point}). Solving $h_{A_C}$ and $h_C$ we obtain new tensors $A_C$ for all $x$ and $y$ and $C$ for all  $x$ and $y$. Using these we update the boundaries using (\ref{nt_truncation_a_l}) and (\ref{nt_truncation_a_r}) obtaining a gauging error. To do so, use either (\ref{nt_truncation_a_l}) or (\ref{nt_truncation_a_r}) to get a set of left (right) canonical tensors from the $A_C$ and $C$ together with the respective gauging errors. Next find the right (left) canonical representation of these tensors as in Ref.~\cite{Vanderstraeten2019a} Algorithm 2. The left and right canonical tensors define the boundaries for the next iteration. Next, start again from the beginning until the gauging error is smaller than a desired threshold $\epsilon$. The algorithm is shown with more structure in Algorithms \ref{p_environment_algorithms} and \ref{p_pvumps_algorithm}. Note, instead of using the gauging error one might as well use the singular values of the $C$ tensors as convergence criterion. Finally, instead of the parallel update just defined one could as well use a sequential update in the $x$ direction (cf. Appendix \ref{appendix:sequential_update}).

\section{Test cases and benchmarks}

\begin{algorithm*}
\SetAlgoVlined
\DontPrintSemicolon
\SetInd{1em}{1em} 

\KwData{MPO $M=(M_{x,y})_{x,y}$ with $(n_x, n_y)$ shaped unit cell; desired accuracy $\epsilon_{prec}$}
\KwResult{Array containing the data of $n_y$ MPS corresponding to the boundaries $(\ket{\psi(A_y)})_y$}
\;
$\left(\ket{\psi(A_y)}\right)_y$ $\longleftarrow$ initialize MPS array\;
$(L_{x,y})_{x,y},\;\;(R_{x,y})_{x,y}\longleftarrow$ update environments from $\{\left(\ket{\psi(A_y)}\right)_y, M\}$  calling \ENvs\;
$\epsilon_{trunc} \longleftarrow 1$\;
$(\epsilon_{trunc, y})_y \longleftarrow (1)_y$\;
\While{$\epsilon_{trunc} > \epsilon_{prec}$}{
$\left(A_C^{x,y}\right)_{x,y} \longleftarrow$ solve $h_{A_C}$ using an iterative solver calling \PAPAC\;

$\left(C^{x,y}\right)_{x,y} \longleftarrow$ solve $h_{C}$ using an iterative solver calling \PAPC\;

\For{$y \in \{0,\dots, n_y\}$}{
$\{A_{L/R}^{x,y}\}_{x}, \epsilon_{trunc, y} \longleftarrow \{(A_C^{x,y})_x, (C^{x,y})_x\}$ following 
(\ref{nt_truncation_a_l}) or (\ref{nt_truncation_a_r})\;
$\ket{\psi(A_y)} \longleftarrow \{A_{L/R}^{x,y}\}_{x}$ by finding the mixed gauge (similar to Algorithm 2 in Ref.~\cite{Vanderstraeten2019a})\;	
}		
$\epsilon_{gauge} \longleftarrow \text{max}\{\epsilon_{gauge, y}\}$\;
$(L_{x,y})_{x,y},\;\;(R_{x,y})_{x,y}\longleftarrow$ update environments from $\{\left(\ket{\psi(A_y)}\right)_y, M\}$  calling \ENvs\;

}

\caption{Parallel implementation of the VUMPS algorithm for non-trivial unit cells.}
\label{p_pvumps_algorithm}
\end{algorithm*}

\subsection{Classical anti-ferromagnetic Ising model}

\begin{figure}
\includegraphics[width=\linewidth]{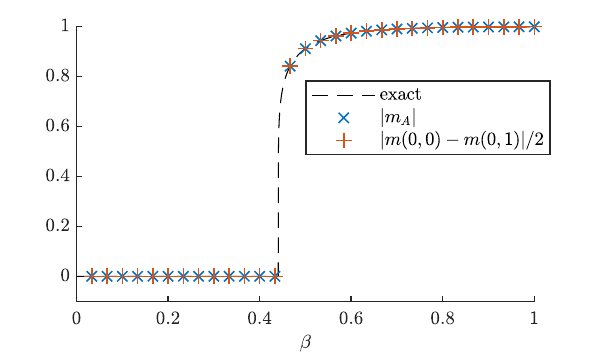}
\caption{Different measures of the magnetization of the anti-ferromagnetic Ising model in the unit cell in the absence of a magnetic field compared to exact results (dashed line) for $\chi=20$. Absolute value of the magnetization at site $(0,0)$ (crosses) and half distance of magnetization between neighbouring sites (pluses).}
\label{fig:af_ising_beta_scaling}
\end{figure}

\begin{figure}
\includegraphics[width=\linewidth]{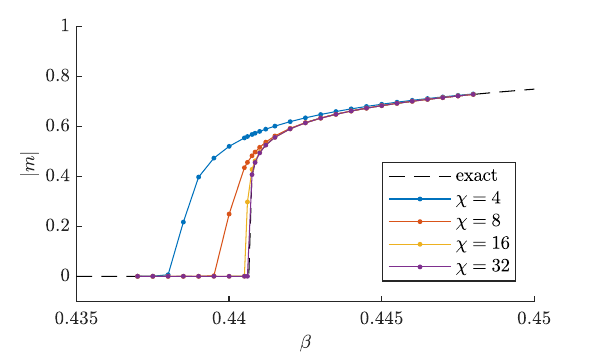}
\caption{Zoom in to the phase transition between the disordered and ordered phase for various values of $\chi=4, 8,16$ and $32$ highlighting the bond dimension dependence of the accuracy of the algorithm. Noteably, for higher bond dimensions the difference between the numerical and exact results around the critical point is no longer visible in the present resolution.}
\label{fig:af_ising_beta_scaling_detailed}
\end{figure}

\begin{figure}
\includegraphics[width=\linewidth]{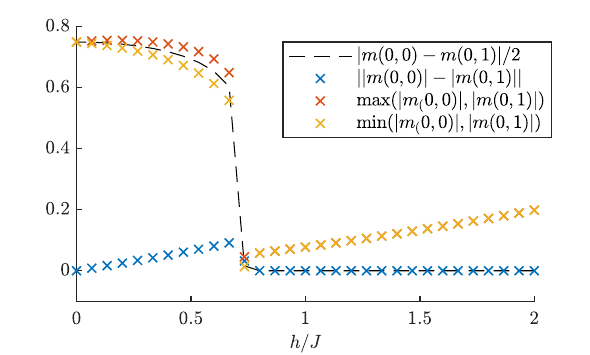}
\caption{Magnetic phase transition at $\beta=0.45$ from the anti-ferromagnetic phase to the magnetically ordered ferromagnetic phase. Various measures are shown such as the half distance of the magnetization between neighbouring sites (dashed line), distance of absolute values of magnetizations between neighbouring sites (blue crosses), maximum (red crosses) and minimum (yellow crosses) of the absolute values of the magnetization at sites $(0,0)$ and $(0,1)$, respectively.}
\label{fig:af_ising_h_scaling}
\end{figure}

To test the multi-site VUMPS algorithm we investigate the classical anti-ferromagnetic Ising model on a square lattice with a magnetic field. This is given by the partition function
\begin{align} \label{af_ising_Z}
&\mcal{Z}(\beta, J, h, N) = \sum_{s\in \{\pm1\}^{\times N}}\exp(-\beta\mcal{H}(s, J, h)),\\
&\mcal{H}(s, J, h) = J\sum_{\langle i,j\rangle} s_is_j + h \sum_i s_i,
\end{align}
where $\langle i,j\rangle$ denotes the sum over nearest neighbours on the square lattice. We can write this partition function as a two-dimensional tensor network on the square lattice from the tensor

\begin{equation} 
\begin{gathered}
\includegraphics[valign=c]{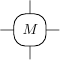}
= 
\includegraphics[valign=c]{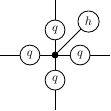}					
\end{gathered}
\end{equation}
where

\begin{equation}
\begin{gathered}
\includegraphics[valign=c]{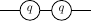}			
= 
\mqty(e^{-\beta J} & e^{\beta J} \\ e^{\beta J} & e^{-\beta J})
\end{gathered}
\end{equation}
and

\begin{equation}
\begin{gathered}
\includegraphics[valign=c]{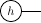}			
= 
\mqty(e^{\beta h}  \\  e^{-\beta h})
\end{gathered}\; .
\end{equation}
Although the partition function of the Ising anti-ferromagnet can be represented by a single tensor $M$, it actually has a non-trivial unit cell. This can be easily seen as follows. The partition function of the anti-ferromagnet equals that of the Ising ferromagnet after a sub-lattice rotation flipping every second spin in the lattice. This transformation has a $2\times2$ unit cell. Also, this sub-lattice rotation would map a homogenous magnetic field to a staggered magnetic field highlighting this unit cell. This actually reflects the fact that the anti-ferromagnetic partition function has a vanishing norm if restricted to odd lattices. Hence, the thermodynamic limit is only well defined if restricting to even lattices. This behaviour can be nicely observed using  our algorithm. So in a sense, the anti-ferromagnet is a ferromagnet with a non-trivial unit cell.

Fixing $J=1$ we can map the physics of Eq.~\eqref{af_ising_Z}
into the plane spanned by $\beta \geq 0$ and $h\in\mathbb{R}$. For $h=0$ we can compute the magnetization within the unit cell and compare to exact results. As for $h=0$ the anti-ferromagnet equals the ferromagnet with every second site flipped in the $z$ basis, we expect a staggered magnetization in the unit cell with the same strength as for the ferromagnet. In Fig.~\ref{fig:af_ising_beta_scaling} we show our results obtained with algorithm \ref{pvumps_algorithm} for bond dimension $\chi=20$. We compare the absolute value of the magnetization of the first site in the unit cell $(x,y)=(0,0)$ with the exact results. Moreover, we compare to the half distance between consecutive sites in the unit cell $\abs{m(0,0)-m(0,1)}/2$ where $m(x,y)$ is the magnetization at the unit cell site $(x,y)$. We find perfect agreement for all quantities up to machine precision.

\begin{figure}  
\includegraphics[width=\linewidth]{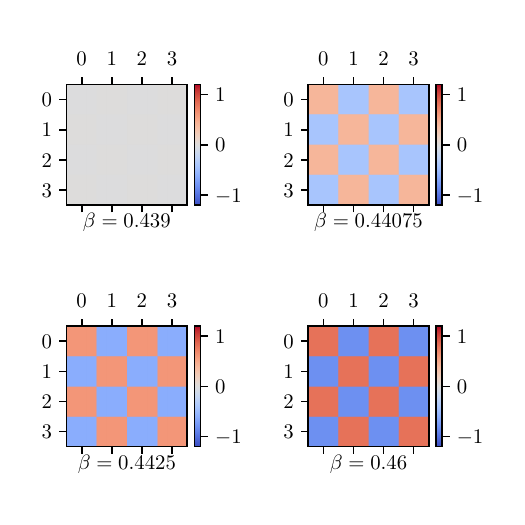}
\caption{Heatmap visualizing the emergence of the anti-ferromagnetic order within the unit cell through the thermal phase transition $\beta=0.439$ (top left), $\beta=0.44075$ (top right), $\beta=0.4425$ (bottom left), $\beta=0.46$ (bottom right) for $\chi=44$.}
\label{fig:af_ising_beta_scaling_uc}
\end{figure}

We have investigated the phase transition around $\beta_c\approx 0.44068$ in more detail in Fig.~\ref{fig:af_ising_beta_scaling_detailed}. We have computed $\abs{m(0,0)}$ for various bond dimensions $\chi=12, 16, 20, 28, 36, 44$. Only very close to the critical point our results start to deviate from the exact values depending on the bond dimension, reflecting the fact that low-$\chi$ MPS cannot describe critical states accurately.
In order to further illustrate our results, in Fig.~\ref{fig:af_ising_beta_scaling_uc} we show the magnetization in the unit cell depending on $\beta$ for $h=0$. One can nicely see the emergence of the staggered magnetization through the phase transition.

\begin{figure}
\includegraphics[width=\linewidth]{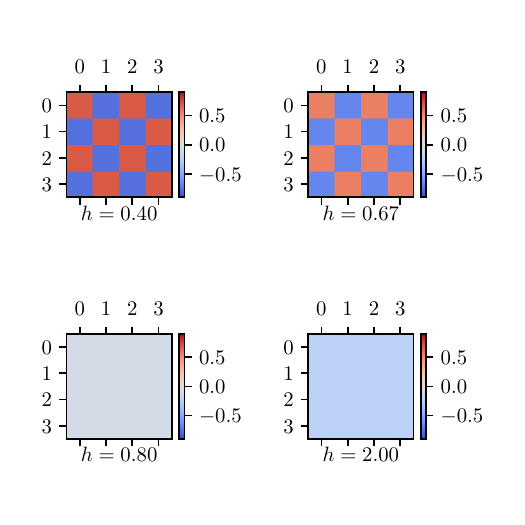}
\caption{Heatmap visualizing the magnetization within the unit cell through the magnetic phase transition for magnetic fields $h=0.4$ (top left), $h=0.6667$ (top right), $h=0.8$ (bottom left) and $h=2$ (bottom right) at $\chi=20$.}
\label{fig:af_ising_h_scaling_uc}
\end{figure}

Next, we have investigated the transition from a staggered magnetized phase to an ordered magnetized phase at $\beta=0.45$ and along $h\in[0, 2]$. In Fig.~\ref{fig:af_ising_h_scaling} we show our results for various quantities obtained from the local magnetizations $m(x,y)$. One can nicely see the linear dependence of the difference between the absolute values of the local magnetizations of consecutive sites until the phase transition around $h_c(\beta=0.45)\approx 0.74$, where it drops to zero and the local magnetizations start to point in the same direction with the same values. In Fig.~\ref{fig:af_ising_h_scaling_uc} we  have visualized this process by plotting the local magnetizations in the unit cell for four consecutive values of $h$ at $\beta=0.45$, nicely showing the transition from an anti-ferromagnetic to a ferromagnetic order.

\subsection{Interacting dimers on the square lattice}

As a second benchmark, we consider a model of interacting dimers on the square lattice  \cite{Alet2005,Alet2006}. This model starts from the well-known dimer-counting problem, which was solved exactly in the early sixties  \cite{Kasteleyn1961, Temperley1961, Fisher1961}. This problem can be extended to a statistical mechanics model, where the allowed configurations are determined by all dimer coverings and we can associate an energy to each dimer configuration. A natural choice for the energy $E_c$ of a given configuration is the number of parallel dimers, 
\begin{equation}
E_c = - \sum_p N_p(c),
\end{equation}
where the sum is over all plaquettes and $N_p(c)$ is either one or zero (depending on whether there are two parallel dimers on plaquette $p$ for the configuration $c$). We include a minus sign in order to favour parallel dimer configurations. The partition function is obtained by summing the Boltzmann weights over all configurations
\begin{equation}\label{partition_function_dimers}
\mathcal{Z} = \sum_c \mathrm{e}^{ \beta \sum_p N_p(c) }.
\end{equation}
with $\beta=1/T$ the inverse temperature. This model has been shown  \cite{Alet2005, Alet2006} to exhibit rotational and translational symmetry breaking in a so-called columnar phase, quantified by the order parameter
\begin{equation}
D = \frac{1}{\mathcal{Z}} \sum_c \mathrm{e}^{ \beta \sum_p N_p(c) } \frac{1}{N} \left( \sum_{l\in\mathcal{V}} M_l(c) - \sum_{l\in\mathcal{H}} M_l(c) \right),
\end{equation}
where $M_l(c)$ is one if there is a dimer on the link $l$ and zero otherwise, and $\mathcal{V}$ and $\mathcal{H}$ denote all vertical resp. horizontal links. On the other hand, in the high-temperature limit ($T\rightarrow\infty$) the partition function reduces to an unweighted sum over all allowed dimer configurations, the model is exactly solvable  \cite{Kasteleyn1961, Temperley1961, Fisher1961} and exhibits critical correlations  \cite{Fisher1963} without any spatial symmetry breaking. A phase transition occurs between these two phases around temperature $T_c=0.65$  \cite{Alet2006, Li2014}.
\par We can represent the partition function of the interacting dimer model by a tensor network with a bipartite sub-lattice structure. Indeed, if we choose the tensors $A$ and $B$ as

\begin{equation}
\begin{gathered}
\includegraphics[valign=c]{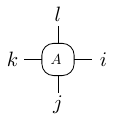}	
= \left\{
\mqty{\,1, \quad & i=j+1=k+2=l+3 \\ 
\,0, \quad & \text{else}} \right. 
\end{gathered}
\end{equation}
respectively

\begin{equation}
\begin{gathered}
\includegraphics[valign=c]{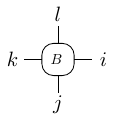}	
= \left\{
\mqty{\,1, \quad & i=j-1=k-2=l-3 \\ 
\,0, \quad & \text{else}} \right. 
\end{gathered}
\end{equation}
and define the matrix $t$ as

\begin{equation}
\begin{gathered}
\includegraphics[valign=c]{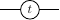}			
= 
\mqty(1 & 0 & 0 & 0 \\ 0 & e^{\beta/2} & 1 & 1 \\ 0 & 1 & 1 & 1 \\ 0 & 1 & 1 & e^{\beta/2} )
\end{gathered}
\end{equation}
we can define 

\begin{equation} 
\begin{gathered}
\includegraphics[valign=c]{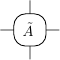}
= 
\includegraphics[valign=c]{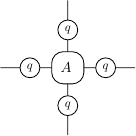}					
\end{gathered}
\end{equation}
with $q^2=t$, and the analogical definition for $\tilde{B}$. Then the partition function Eq.~\eqref{partition_function_dimers} is given by

\begin{equation}
\begin{gathered}
\mcal{Z} =
\includegraphics[valign=c]{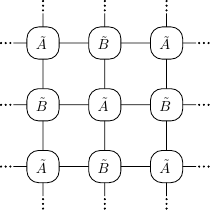}			
\end{gathered}.
\end{equation}

\begin{figure}
\includegraphics[width=\columnwidth]{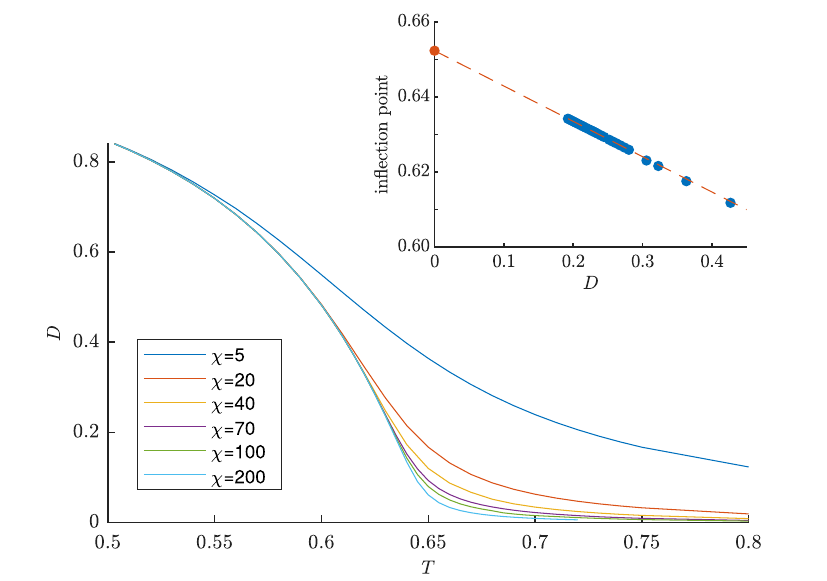}
\caption{The order parameter $D$ characterizing the columnar phase in the interacting dimer model as a function of temperature for different values of the bond dimension. In the inset we perform an extrapolation of the inflection point as a function of the bond dimension $D$, yielding an estimate for the critical point $T_c\approx0.6523$.}
\label{fig:order}
\end{figure}

\begin{figure*} \centering
\subfigure[$\quad T=0.2$]{\includegraphics[width=0.9\columnwidth]{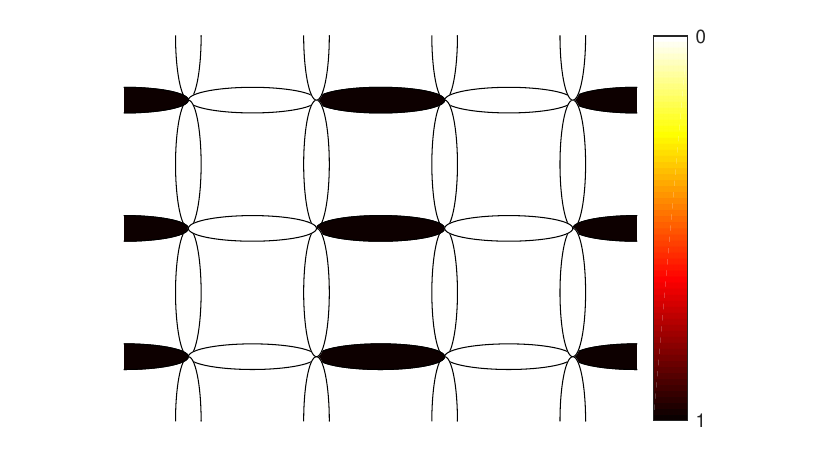}}
\subfigure[$\quad T=0.6$]{\includegraphics[width=0.9\columnwidth]{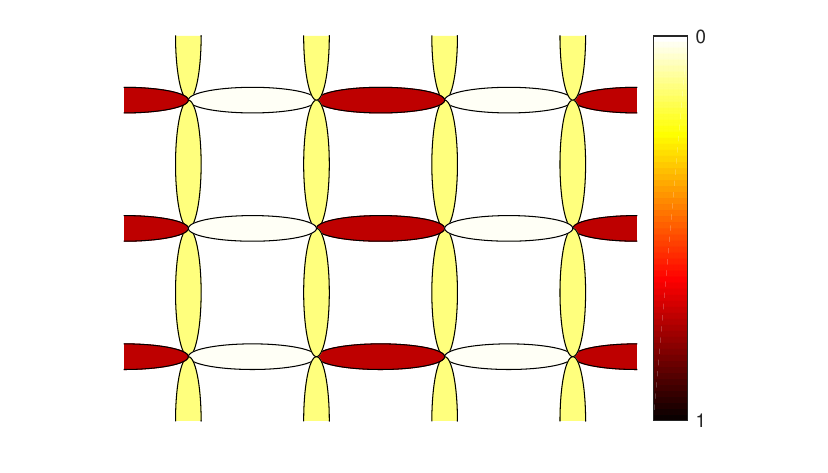}}\\
\subfigure[$\quad T=0.7$]{\includegraphics[width=0.9\columnwidth]{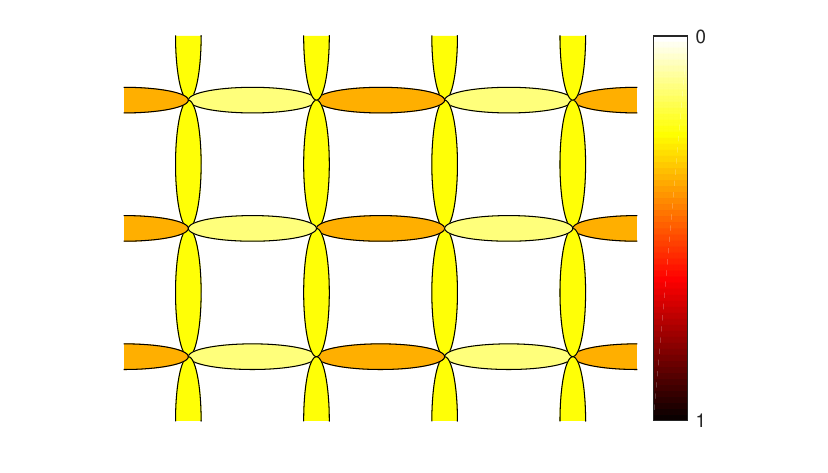}}
\subfigure[$\quad T=0.8$]{\includegraphics[width=0.9\columnwidth]{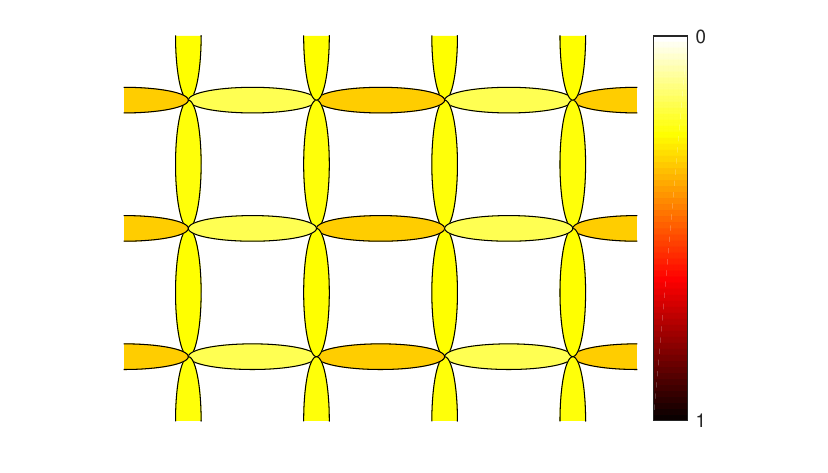}}
\caption{Probability (color scale) of finding a dimer on different links in the lattice for different temperatures, as computed by the VUMPS algorithm with bond dimension $\chi=70$.}
\label{fig:dimers}
\end{figure*}

This construction is seen to give the correct partition function by realizing that (i) a `1' on a link denotes the presence of a dimer in that configuration, (ii) the tensors $A$ and $B$ guarantee that exactly one dimer is on each site, (iii) the $e^{\beta/2}$ factors in the $t$ matrix introduce the correct Boltzmann weight when there are two parallel dimers on a plaquette.
\par Since we have a tensor network with a two-by-two unit cell and this model breaks translational symmetry at low temperatures, we apply the multisite VUMPS algorithm. In Fig.~\ref{fig:dimers} we plot the probabilities of finding a dimer on the different links in the lattice, showing columnar order for low temperatures and a uniform distribution in the high-temperature phase. In Fig.~\ref{fig:order} we show the behaviour of the order parameter $D$ for different values of the bond dimension, showing an increasingly critical form as the bond dimension increases. To determine the critical point we compare the temperature of the inflection point against the order parameter of the inflection point for different values of the bond dimension. We see that the temperature of the inflection point relates linearly with the value of the order parameter at that temperature. In the inset of Fig.~\ref{fig:order} we provide an extrapolation of the critical point $T_c=0.6523\pm0.0001$, which agrees with and improves upon known values in the literature.

\subsection{Doped RVB state}

For our third benchmark we start from the resonating valence bond (RVB) state in the square lattice, which can be represented as a translation-invariant PEPS from a local tensor $A_{u,r,d,l}^s$ with explicit $\mathrm{SU}(2)$ invariance  \cite{Schuch2012}. Following the framework of symmetric tensor networks, we can label the non-zero blocks in the 
tensor with $\mathrm{SU}(2)$ irreducible representations; for the nearest-neighbour RVB we have only four non-zero blocks,
\begin{align}
&A^{\frac{1}{2}}_{\frac{1}{2},0,0,0} = 1, \quad A^{\frac{1}{2}}_{0,\frac{1}{2},0,0} 
= 1, \nonumber \\ & A^{\frac{1}{2}}_{0,0,\frac{1}{2},0} = 1, \quad A^{\frac{1}{2}}_{0,0,0,\frac{1}{2}} = 1.
\end{align}
This state is known  \cite{Schuch2012} to be in a critical $(2+0)$-dimensional Coulomb phase and has power-law decaying dimer-dimer correlations. As a result, the contraction typically requires a large bond dimension for accurate results.
\par We now modify the RVB state by introducing holes on one of the two sub-lattices. This `doping' is performed by introducing a second PEPS tensor $B$ with non-zero blocks
\begin{align}
& B^0_{0,0,0,0} = \lambda_1 ,\\
& B^0_{\frac{1}{2},\frac{1}{2},0,0}=\lambda_2, \quad B^0_{0,\frac{1}{2},\frac{1}{2},0} = \lambda_2, \nonumber \\ & B^0_{0,0,\frac{1}{2},\frac{1}{2}} = \lambda_2, \quad B^0_{\frac{1}{2},0,0,\frac{1}{2}} = \lambda_2, \\
& B^0_{0,\frac{1}{2},0,\frac{1}{2}} = \lambda_3, \quad B^0_{\frac{1}{2},0,\frac{1}{2},0} = \lambda_3.
\end{align}
The PEPS is then built up from the PEPS tensors $A_1=A$ and $A_2=A+B$ as

\begin{equation}
\begin{gathered}
\ket{\Psi(A_1,A_2)} =
\includegraphics[valign=c]{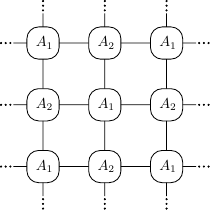}			
\end{gathered},
\end{equation}
where the above blocks in the tensor $B$ are chosen such that the state is rotation invariant. This state is similar to previous doping constructions, where a doping of the RVB state by unpaired spins  \cite{Poilblanc2014} and fermionic holes  \cite{Poilblanc2014b} has been implemented in a translation-invariant way. Also, in contrast to the latter, we do not invoke any fermionic degrees of freedom in our state, so we effectively dope the state with bosonic holes on every second site.

\begin{figure}
\includegraphics[width=\columnwidth]{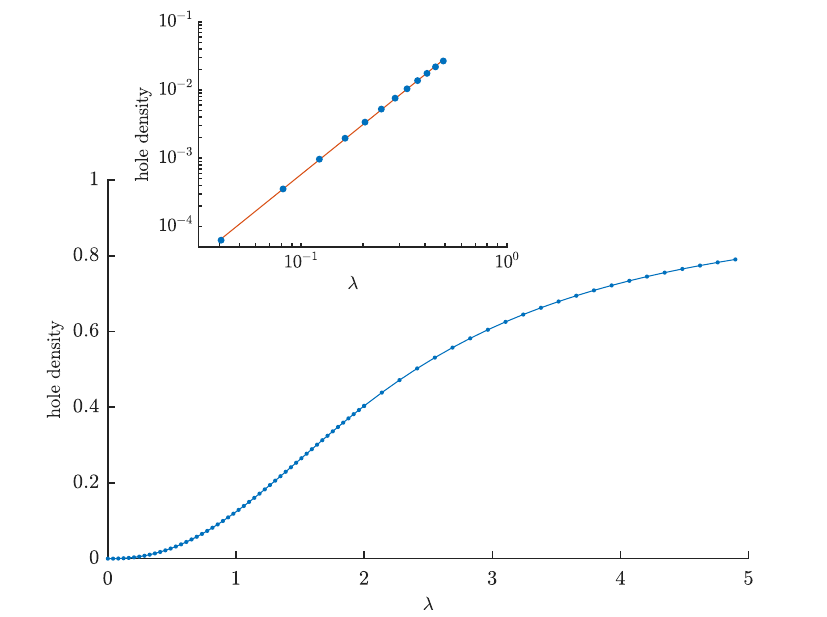}
\caption{The hole density of the doped RVB state as a function of a single parameter $\lambda$. We have used a $\mathrm{SU(2)}$ symmetric version of the multisite VUMPS algorithm with total bond dimensions ranging from $92$ to $1414$. For small $\lambda$ we see a power law behaviour, as illustrated by the inset log-log plot (we find the exponent $2.43\pm0.02$).}
\label{fig:rvb}
\end{figure}

\par Since this PEPS is non-translation-invariant by construction, we use the multi-site VUMPS algorithm for contracting it and computing observables. As an illustration, we fix the above three parameters to be the same ($\lambda_1=\lambda_2=\lambda_3=\lambda$) and we compute the hole density (per site) as a function of the parameter $\lambda$. Because the undoped RVB state is critical, a large bond dimension for the boundaries is needed and we work with explicit $\mathrm{SU}(2)$ invariant tensors in each step of the VUMPS algorithm. In Fig.~\ref{fig:rvb} we plot the result, showing a slow onset that follows a power-law, as is illustrated by the inset log-log plot followed by a slow saturation to $1$ as $\lambda$ grows bigger, as is to be expected.

\section{Conclusion and outlook}

In this work, we have presented a generalization of the VUMPS algorithm for contracting two-dimensional tensor networks to the case of non-trivial unit cells. The algorithm inherits the salient features of variational uniform MPS methods including optimality guarantees and high rates of convergence, while the computational effort scales only linearly with the size of the unit cell. We have benchmarked the contraction method on the square lattice antiferromagnetic Ising model, the phase transition in the square-lattice interacting dimer model and on the doped RVB PEPS.
We expect this algorithm to be important in the simulation of quantum spin systems in two-dimensions with PEPS, wherever larger unit cells are required for the representation of the ground state, as well as for the simulation of classical two-dimensional systems where a unit cell appears either in the representation of the partition function or due to symmetry breaking.
We hope that the emulation technique of larger unit cells presented in this work inspires further generalizations of known tensor network methods to larger unit cells.

\section*{Acknowledgments}
AN thanks A. Bauer for stimulating discussions and proofreading and is grateful to the ObstOffice for a fruitful and juicy environment. This work is supported by the Research Foundation Flanders, ERC grants QUTE (No. 647905), 
the DFG (CRC 183, EI 519/15-1), the Templeton Foundation, and EU Grant SIQS.

\bibliography{bibliography.bib}

\appendix
\newpage

\section{Derivation of the generalized fixed point equations}
In this appendix, we are going to give an alternative to the derivation of the optimality criterion of the VUMPS algorithm and thereby generalize it to non trivial unit cells. In Section \ref{topo_argument} we will generalize the reasoning from Ref.~\cite{ZaunerStauber2018} to non-hermitian transfer operators. In Section \ref{variational_argument} we will further generalize to non-trivial unit cells. An alternative derivation of the fixed point equations for a non-trivial unit cell using the emulation of non-trivial unit cell tensor networks with trivial unit cell tensor networks which is based on the single site VUMPS algorithm with non-hermitian transfer operators is given in Appendix \ref{appendix:non_trivial}.

\subsection{VUMPS for non-hermitian transfer matrices}\label{topo_argument}
The VUMPS algorithm \cite{ZaunerStauber2018} is a scheme to optimize the equation
\begin{equation}\label{optimality_criterion}
\mathcal{P}_A(M-\lambda)\ket{\psi(A)}=0
\end{equation}
via iterating over local tensor equations. Here $A$ is the tensor parametrizing the state vector $\ket{\psi(A)}$, $M$ is the MPO for which we wish to find an approximation of the boundary, $\lambda$ is the corresponding maximal eigenvalue and $\mathcal{P}_A$ is the projector onto the tangent space of the MPS manifold at the point parametrized by $A$ as in Eq.~\eqref{tangent_space_projector}. Eq.~\eqref{optimality_criterion} can be shown to be the variational optimum approximating the boundary of the tensor network parametrized by $M$ over the MPS manifold for any given fixed bond dimension given that the MPO transfer matrix defined through $M$ is hermitian \cite{ZaunerStauber2018}.
In the following, we are going to show that this holds true also if the MPO transfer matrix defined through $M$  ceases to be hermitian but is gapped, in the sense that the the absolute values of the eigenvalues of the MPO transfer matrix still sustain a gap stable in the asymptotic limit.
This situation can reflect a classical partition function of a non-critical Ising model or the double layer sandwich of the toric code PEPS.

For the contraction of the tensor network, we are interested in its boundary which is the maximal eigenvector of the MPO transfer matrix. More precisely, the boundary $\ket{\phi}$ is the eigenvector corresponding to the largest eigenvalue of $M$ in absolute value
\begin{equation}\label{topo_argument_fp_equ}
M\ket{\phi}\propto\ket{\phi}
\end{equation}
This, in turn, corresponds to the resulting state upon an infinite application of the MPO transfer matrix on a suitably chosen random initial state. Moreover, for topological fixed point models, the maximal eigenvectors correspond to the topological boundaries of the model. Those are the states that live on the boundary if one computed observables of this model on a manifold with a boundary.

Because the MPO is gapped in the above sense, the boundary will feature exponentially decaying correlations. The exponentially decaying correlations of the one dimensional boundary, however, imply that we can faithfully approximate it locally by an MPS. More precisely, the reduced density matrix of the many body state with exponential correlations on segments of length $k$ is $\epsilon$-close to an MPS with bond dimension scaling as $\bigO(k/\epsilon^3)$ \cite{Brandao2015, schuch2017, dalzell2019, huang2019}. So in this sense, being interested only in $k$-local properties we can write the state explicitly as an MPS with an error smaller than~$\epsilon$. 

Hence, for large enough $k$ and at a sufficiently large bond dimension $\chi$ such that the corresponding $(k,\epsilon)$ approximation error is below machine precision, we are practically not able to computationally distinguish the MPS $\ket{\psi(A)}$ from the true boundary state vector $\ket{\phi}$. Hence, we argue that we can apply the fixed point equation Eq.~\eqref{topo_argument_fp_equ} to $\ket{\psi(A)}$
\begin{equation}\label{topo_argument_approx}
M\ket{\psi(A)} \approx \lambda\ket{\psi(A)}
\end{equation}
where the approximation error due to the application of $M$ might increase from $\epsilon$ to some $\epsilon'>\epsilon$, though the increase will be in a controlled way given that
$M$ is gapped. 
This can be understood as follows. The state vector $\ket{\psi(A)}$ is $(k,\epsilon)$-close to $\ket{\phi}$. Additionally, the application of $M$ exponentially suppresses the overlap orthogonal to its maximal eigenstate. Hence, the normalized $M\ket{\psi(A)}$ is closer to $\ket{\phi}$ than $\ket{\psi(A)}$ is. Therefore, the normalized $M\ket{\psi(A)}$ will be $(k',\epsilon'/2)$ close to $\ket{\phi}$, up to boundary effects (at the boundary of the local patch of size $k$) decaying exponentially in the depth into the region, leading to a modified $\epsilon'/2$, and hence at most $(k', \epsilon')$-far from $\ket{\psi(A)}$. Hence, if $k$ and $\chi$ are chosen sufficiently large, the resulting $\epsilon'>0 $ will still be below machine precision, while $k'$ will still be sufficiently large.

As we are working over the manifold of MPS, it is natural to use the tangent space projector in order to reduce the fixed point Eq.~\eqref{topo_argument_approx} from a many body Hilbert space equation to a finite equation. The resulting equation still resolves the overlap of $M\ket{\psi(A)}$ with the full manifold of MPS whilst only neglecting states that cannot be captured by the manifold of MPS. Hence, we obtain Eq.~\eqref{optimality_criterion}. Similarly as before, one might argue that the projector is not a local operator and hence one might not be able to apply the $(k,\epsilon)$-closeness. However, we can construct a similar argument as for the MPO transfer matrix applied to the MPS in Eq.~\eqref{topo_argument_approx}. For each term in the sum of the projector (cf.~Eq.~\eqref{tangent_space_projector}) applied to the MPS, 
we find that essentially only a finite patch contributes to the projector because the gap in the MPS-MPO-MPS transfer matrix exponentially suppresses the contribution of the tail of the transfer matrix channel. 

It is interesting to note that the above reasoning could be generalized even further to critical systems, at the price of a polynomial rather than a linear dependence on the bond dimension \cite{huang2019}.
Also, for critical systems, the application of $M$ does not exponentially suppress the overlap orthogonal to the boundary but only in a weaker sense.
To conclude, we find that the optimality criterion the VUMPS algorithm optimizes can also be understood for non-hermitian gapped MPOs.

\subsection{Non-trivial unit cells: A variational argument}\label{variational_argument}

As explained in Section \ref{section:non_trivial_vumps}, in order to find the boundary of a tensor network with a non-trivial unit cell we need to find a $\ket{\psi_1}$ and $\lambda$ that satisfy $M_n\dots M_2M_1\ket{\psi_1}=\lambda\ket{\psi_1}$, where the $M_i$ are MPOs or MPO-like objects defining the tensor network. We do this by assuming $\ket{\psi_1}$ is well captured by an MPS $\ket{\psi(A_1)}$, which may be motivated through arguments similar as in Appendix \ref{topo_argument}. As explained before, this is equivalent to looking for a set of MPSs $\ket{\psi(A_i)}$ and scalars $\gamma_i$ satisfying
\begin{equation}\label{variational_boundary_equs}
\begin{aligned}
M_1\ket{\psi(A_1)}&\approx\gamma_1\ket{\psi(A_2)},\\
M_2\ket{\psi(A_2)}&\approx\gamma_2\ket{\psi(A_3)},\\
&\dots\\
M_n\ket{\psi(A_n)}&\approx\gamma_n\ket{\psi(A_1)},
\end{aligned}
\end{equation}
with $\prod_i\gamma_i=\lambda$. We use the approximate sign in order to emphasize that $\ket{\psi(A_{i+1})}$ approximates $M_i\ket{\psi(A_i)}$ in some sort of optimal way. In this text we choose the meaning of this optimality to be variational in the sense of the fidelity per site in the thermodynamic limit. In particular,
\begin{equation}
\lim\limits_{N\to\infty}|\mel{\psi^N(A_{i+1})}{M_i}{\psi^N(A_i)}|^{1/N}
\end{equation} 
should be maximal with respect to $A_{i+1}$.

Having thus translated the problem to variational MPS \cite{Vanderstraeten2019a} we can formulate a first naive algorithm to find the boundary.
Start with some initial MPS $\ket*{\psi(A_1^{(0)})}$ and consecutively apply all the $M_i$ each time followed by variationally approximating the new MPS by maximizing the fidelity per site by another MPS of some given bond dimension. Repeat that procedure until convergence is reached. Clearly, this power method will eventually converge to the fixed point characterized by the Eq.~\eqref{variational_boundary_equs}. 

Similarly to the derivation of the VUMPS algorithm \cite{ZaunerStauber2018} we find the optimality criterion of maximizing the fidelity per site to be equivalent to the vanishing of the tangent space projector on the fixed point equation. In particular, for the variationally optimal solution $A_i$ to Eq.~\eqref{variational_boundary_equs} it must hold 
\begin{equation}
\mathcal{P}_{A_{i+1}}(M_i\ket{\psi(A_i)}-\gamma_i\ket{\psi(A_{i+1})})=0
\label{tangent_space_projector_eq}
\end{equation}
where $\mathcal{P}_{A}$ represents the MPS-tangent space projector at the state parametrized by $A$. In this particular case the tangent space criterion is equivalent to maximizing the fidelity per site. Note, however, that the tangent space criterion need not imply a variational principle in general. 

Instead of the just described power method however we go one step further: we write down all the fixed point equations Eq.~\eqref{variational_boundary_equs} in the projected form of Eq.~\eqref{tangent_space_projector_eq} at once, and call this our new, global fixed point equation. This can then be solved iteratively in a similar spirit as the original VUMPS algorithm as explained in Section \ref{section:algorithm}. It is easy to see that the set of these global fixed point equations corresponds to \ref{local_AC_eigenvalue_equation} and \ref{local_C_eigenvalue_equation}. Obviously, as Eq.~\eqref{tangent_space_projector_eq} is contained in the global fixed point equation a solution to this global fixed point equation contains the result of the above described power method.

Note that the same chain of reasoning starting from the variational power method and concluding with the VUMPS fixed point iteration can be made for the standard single site VUMPS algorithm. However, for a Hermitean operator $H$ one may also derive the VUMPS equations directly from the more common global  variational principle
\begin{align}
\max_A f(A, \bar{A})=\frac{\mel{\psi(A)}{H}{\psi(A)}}{\braket{\psi(A)}}.
\end{align}
This follows from the fact that the VUMPS condition can be represented as $\partial_{\bar{A}} f(A,\bar A) = 0$, which implies $\partial_A f(A,\bar A) = 0$. Hence, the total derivative of $f$ is zero at the VUMPS fixed point. We can generalize this to two layered transfer matrices $T_1 T_2$ with the special structure that they compose to a Hermitean transfer matrix: $T_2^\dagger=T_1= T$. Define the cost function 
\begin{align}
  f(A_1,\bar{A}_1&,A_2,\bar{A}_2)=\\
  &=\frac{\bra{\psi(A_1)} T^\dagger \ketbra{\psi(A_2)} T \ket{\psi(A_1)}}{\braket{\psi(A_1)}\braket{\psi(A_2)}}.\nonumber
\end{align}
The global projected fixed point equations then correspond $\partial_{A_1}f=\partial_{A_2}f=0$. Moreover, due to the Hermiticity of $T^\dagger T$ the function $f$ is real such that this also implies $\partial_{\bar A_1}f=\partial_{\bar A_2}f=0$. Hence, the multi site VUMPS equations in the special case of a hermitean two layered transfer matrix directly corresponds to a global variational principle similar as for single site VUMPS.

For any number of layers greater than two, this argument can not be repeated. We illustrate this here with three layers. Consider the analogous cost function $f$ defined as follows
\begin{equation}
\frac{\bra{\psi(A_1)} T^\dagger \ketbra{\psi(A_2)} H \ketbra{\psi(A_3)} T \ket{\psi(A_1)}}{\braket{\psi(A_1)}\braket{\psi(A_2)} \braket{\psi(A_3)}}\,.
\end{equation}
where $H$ is chosen Hermitian such that the total transfer matrix $T^\dagger HT$ is Hermitian. The corresponding fixed point equations are still given by $\partial_{\bar{A}_1} f = 0$, $\partial_{\bar{A}_2} f = 0$ and $\partial_{\bar{A}_3} f = 0$. However, neither is $f$ guaranteed to be real, nor are the partial derivatives implied by their adjoint counter part. The fixed point equation thus does not follow from a variational principle. In fact, while the fixed point equation proposed in this work characterizes the actual solution to the boundary equation there is no corresponding global variational principle whose extremum corresponds to the solution of the three boundary equations simultaneously. 

An underlying global variational principle can be very helpful in stabilising the corresponding method: If the iterative VUMPS method on a Hermitian transfer matrix fails to converge, one can resort to a gradient based optimisation scheme on $f$ and still find the fixed point (disregarding local minima). For two dimensional tensor networks with a unit cell that is both wider and taller than two we know of no variational principle that can be used efficiently. Hence, in these cases there is only the method proposed here together with TEBD and CTMRG, neither of which can be guaranteed to converge.

\begin{widetext}
\section{Tensor networks with non-trivial unit cell}\label{appendix:non_trivial}
In this appendix we will formally generalize the VUMPS method to tensor networks that admit a non-trivial unit cell. Instead of arguing from a power method as in Appendix~\ref{variational_argument} we will be using the emulation of non-trivial unit cell tensor networks by trivial unit cell tensor networks. It will be easy to see that the cost scales only linearly rather than an exponential in the unit cell size. In this work, we assume that the lattice on which the tensor network is defined still has the same regularity within the unit cell as on the outer level connecting those unit cells, both having rectangular structure.

\subsection{Emulating tensor networks with a non-trivial unit cell by tensor networks with a trivial unit cell}\label{appendix:non_trivial_tensor_networks}
In order to apply the VUMPS methodology to tensor networks with a non-trivial unit cell we are going to emulate the latter by a tensor network with a trivial unit cell. However, instead of achieving this by merely blocking the tensors within one unit cell -- leading to an exponential overhead in the complexity -- we are going to define a new model with a tensor network with a richer structure leading only to a linear overhead.

To start with, we observe that the data of a two dimensional tensor network on a rectangular lattice is defined by a four leg tensor for any position in the unit cell. In particular, this data can be modeled by a five leg tensor, where the fifth leg carries as a basis the sites of the unit cell. Explicitly, the basis of the fifth leg is given by $\mathbb{Z}_{n_xn_y}\simeq \mathbb{Z}_{n_x}\times\mathbb{Z}_{n_y}$ and we can split this leg into two legs each with basis $\mathbb{Z}_{n_x}$ and $\mathbb{Z}_{n_y}$, respectively. This corresponds to making the $x$ and $y$ dependence of the tensors explicit. Graphically, the defining data is given by the tensor

\begin{equation}
\begin{gathered}
\includegraphics[valign=c]{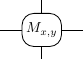}		
=
\includegraphics[valign=c]{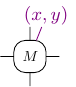}		
=
\includegraphics[valign=c]{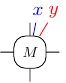}
\end{gathered}
\end{equation}
for all $x,y$ from the unit cell. Starting from this tensor, we use the additional structure

\begin{equation}\label{ttype_nt_mpo}
\begin{gathered}
\includegraphics[valign=c]{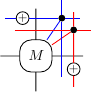}
\end{gathered}
\end{equation}
in order to define the tensor network with trivial unit cell emulating the original tensor network with non-trivial unit cell by connecting them uniformly over a square lattice

\begin{equation}
\begin{gathered}
\includegraphics[valign=c]{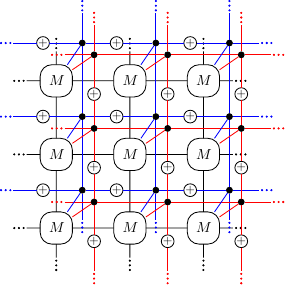}			
\end{gathered} .
\end{equation} 
This new tensor network is the superposition of all unit cell shifts of the original tensor network. Similarly, the red and blue bonds can be seen as defect lines in the two dimensional tensor network corresponding to the symmetry breaking. The original tensor network is then emulated via symmetry breaking into specific unit cell channels and working therein. Moreover, it is easy to see that the new tensor network still carries the information about the unit cell: working on a torus with width $m_x\times m_y$ the tensor network will vanish for any $m_x\neq k_xn_x$ and $m_y\neq k_yn_y$.

In order to efficiently parametrize the boundary of such tensor networks we introduce a generalization of MPS. This is, we generalize the tensor network structure of MPS to a structure related to Eq.~\eqref{ttype_nt_mpo}
\begin{equation}\label{ttype_nt_mps}
\begin{gathered}
\includegraphics[valign=c]{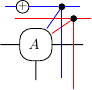}
\end{gathered}.
\end{equation}

This is equivalent to a list of MPS each with a non-trivial unit cell. The substructure of the tensor network of an MPS constructed from tensors as in Eq.~\eqref{ttype_nt_mps} is designed such that we are actually emulating non-trivial unit cell tensor networks when the MPS is placed on the boundary of tensor networks as in Eq.~\eqref{ttype_nt_mpo}.
Clearly, by blocking the indices we re-obtain an MPS however at bond dimension $n_xn_y\chi$  and physical dimension $n_xn_yD$.

For MPS as in Eq.~\eqref{ttype_nt_mps} we can define a gauge in a similar way as in \ref{nt_mps}
\begin{equation}
\begin{gathered}\label{tt_nt_mps}
\ket{\psi(A)} =
\includegraphics[valign=c]{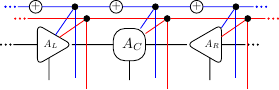}
\end{gathered}
\end{equation}
where the injectivity of the MPS holds in each of the $n_xn_y$ channels encoding the unit cell separately. Also, the state in Eq.~\eqref{tt_nt_mps} is not normalized in the usual sense. Rather, the MPS corresponds to $n_y$ MPS with a non-trivial unit cell in the $x$-direction, each of which being normalized. Similarly, the MPS in Eq.~\eqref{tt_nt_mps} is such that normalization holds in each of the $n_xn_y$ channels separately. In what follows, we will alway refer to \emph{normalized} as normalized in each unit cell channel $(x, y)$ separately. In order to emulate non-trivial unit cell tensor networks via tensors as in Eq.~\eqref{ttype_nt_mpo} and \eqref{ttype_nt_mps} we associate the physical degrees of freedom to the black bonds of the tensor network. The red and blue bonds correspond to the $x$- and $y$-labels of the unit-cell, and can be thought of as a book keeping. 
The set of normalized states parametrized as in Eq.~\eqref{ttype_nt_mps} forms the manifold of non-trivial unit cell MPS with bond dimension $\chi$ and unit cell shape $(n_x, n_y)$ which we denote by $\mathcal{M}_{n_x, n_y, \chi}$. 
Using this tensor network the fixed point equation then reads
\begin{equation}
\label{tt_boundary_equ}
\begin{gathered}
\includegraphics[valign=c]{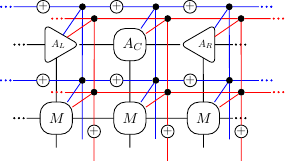}
\approx \gamma
\includegraphics[valign=c]{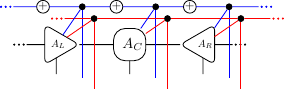}			
\end{gathered}
\end{equation}
Using $+^T=-$ and the fact that the $y$ legs are all connected through delta tensors only we find that Eq.~\eqref{tt_boundary_equ} is equivalent to Eq.~\eqref{pre_MPO_eigenvalue_equation}.
Note that if the MPO in Eq.~\eqref{tt_boundary_equ} is not \emph{normalized} such that it maps the $y$-th normalized boundary MPS to the normalized $y+1$'st boundary MPS we must allow the boundary MPS to be not normalized within the unit cell channels in order to absorb the different scalar factors from the MPOs in each $y$ channel similar to Eq.~\eqref{pvumps_core}.

\subsection{Geometry of MPS with non trivial unit cell tensors}\label{appendix:non_trivial_geometry}
As explained in the previous section we are going to work not on the usual manifold of MPS but on the manifold of MPS with non-trivial unit cell tensors $\mathcal{M}_{n_x,n_y,\chi}$. Therefore, let us re-derive the geometric picture of MPS on the new manifold $\mathcal{M}_{n_x, n_y, \chi}$ similar as in Ref.~\cite{Vanderstraeten2019a}. To this end, we are mainly interested in the tangent space projector on $\mathcal{M}_{n_x, n_y, \chi}$. It is straightforward to compute the tangent vectors $\ket{\phi(B,A)}$ at the point $\ket{\psi(A)}$. They are given by the equal weight superposition of all displacements of defects on $\ket{\psi(A)}$ with a given defect tensor $B$
\begin{equation}
\begin{gathered}\label{nt_tangent_vector}
\ket{\phi(B,A)} = \sum_{x,y,\alpha,s,\beta}B_{x,y,\alpha,s,\beta}\frac{\partial}{\partial A_{x,y,\alpha,s,\beta}}\ket{\psi(A)}=
\sum_{n\in\mathbb{Z}}
\includegraphics[valign=c]{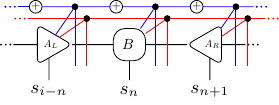}
\end{gathered}
\end{equation}
where $n$ indicates the site on which the defect matrix is placed. Due to the normalization of the MPS manifold the tangent space $\mathcal{T}_A\mathcal{M}_{n_x, n_y, \chi}$ at a point $\ket{\psi(A)}$ must be orthogonal in Hilbert space to the the state parametrized by that point. This, together with the gauge degree of freedom in the MPS eliminates $\chi^2n_xn_y$ degrees of freedom from the original $\chi^2dn_xn_y$ degrees of freedom. This can be addressed directly by requiring that for the tensor $B$ in Eq.~\eqref{nt_tangent_vector} it must hold
\begin{equation}
\begin{gathered}
\includegraphics[valign=c]{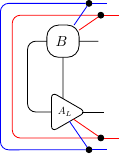}
= 0
\end{gathered}
\end{equation}
and vice versa for the same diagram flipped about the horizontal axis. In particular, $B$ must have support on the 
null space of $A_L$ only. In order to parametrize such tensors $B$ we can define the tensor $V_L$ similar to Ref.~\cite{Vanderstraeten2019a} to be an isometry on on the null space of $A_L$ at each $(x,y)$. Hence, for $V_L$ it must hold
\begin{equation}
\begin{gathered}
\includegraphics[valign=c]{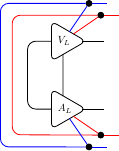}
= 0
\end{gathered}
\end{equation}
and		
\begin{equation}
\begin{gathered}
\includegraphics[valign=c]{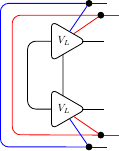}
= 
\includegraphics[valign=c]{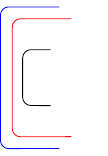}
\end{gathered} .
\end{equation}
As the null space of each of the $n_xn_y$ $A_L$ is $\chi(d-1)$ dimensional, the right bond of $V_L$ at $(n_x, n_y)$ has dimension $\chi(d-1)$. Finally, we can write the parametrization of $B$ using the $n_xn_y$ matrices $X_{n_x,n_y}$ with dimensions $\chi(d-1)\times \chi$ fixing the free degrees of freedom of $B$ as
\begin{equation}
\begin{gathered}
\includegraphics[valign=c]{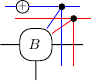}
=
\includegraphics[valign=c]{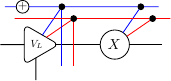}		
\end{gathered} .
\end{equation}
Similarly as in Ref.~\cite{Vanderstraeten2019a}, 
we can rewrite the projector onto the null space as 
\begin{equation}
\begin{gathered}
\includegraphics[valign=c]{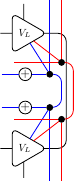}	
= 
\includegraphics[valign=c]{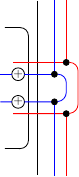}
-
\includegraphics[valign=c]{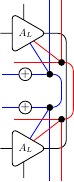}	
\end{gathered} .
\end{equation}
This is, we can straightforwardly use $V_L$ in order to to write the tangent space projector $\mathcal{P}_A$ at point $\ket{\psi(A)}$ acting on states of the form as in Eq.~\eqref{tt_nt_mps} explicitly as 
\begin{equation}
\begin{gathered}
\label{product_tangent_space_projector}
\mathcal{P}_{A} = \sum_{n\in\mathbb{Z}}
\includegraphics[valign=c]{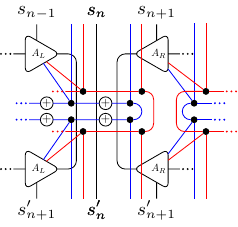}	
-
\includegraphics[valign=c]{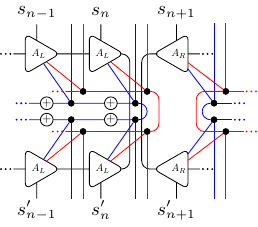}	
\end{gathered}
\end{equation}

\subsection{Multisite VUMPS in explicit notation}
In order to derive the local VUMPS equations for the generalized tensor network we can combine the above Sections \ref{topo_argument}, \ref{appendix:non_trivial_tensor_networks} and \ref{appendix:non_trivial_geometry}. Similar as for standard VUMPS we define the tensors $A_C'$ and $C'$ as
\begin{equation}
\begin{gathered}
\frac{1}{\mathcal{N}}
\includegraphics[valign=c]{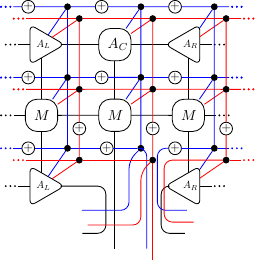}	
=	 
\includegraphics[valign=c]{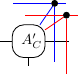}			
\end{gathered}
\end{equation}
and

\begin{equation}
\begin{gathered}
\frac{1}{\mathcal{N}'}
\includegraphics[valign=c]{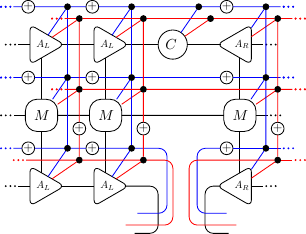}	
=	
\includegraphics[valign=c]{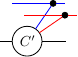}			
\end{gathered}
\end{equation}
where the diverging $\mathcal{N}$ and $\mathcal{N}'$ denote the normalizations counteracting the non-normalized transfer matrices as before. It is worth noticing that due to the substructure in the tensor network spanned by the $x$- and $y$-legs respectively the only non-vanishing elements on the left hand side are exactly those where all $x$-index configurations (respectively $y$-index configurations) match. This means, that we are free to absorb all open $x$-legs ($y$-legs) on the left hand side into one delta tensor with one open leg whilst dropping the delta tensor on the right hand side splitting the $x$-leg ($y$-leg) into three legs, respectively two legs. In other words: The tensor network structure involving the delta tensors on the left hand side is automatically fulfilled by the tensor network on the right hand side (independent of the $M$, $C$ and $A_{R/L}$ tensors).

Using the tensors $A_C'$ and $C'$ it is easy to see that the optimality criterion Eq.~\eqref{optimality_criterion} where the tangent space projector of the form of Eq.~\eqref{product_tangent_space_projector} is acting on the boundary equation in the form of \eqref{tt_boundary_equ} is equivalent to 
\begin{equation}
\label{tt_center_definition}
\begin{gathered}
\includegraphics[valign=c]{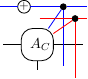}
=
\includegraphics[valign=c]{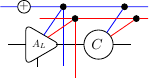}
=
\includegraphics[valign=c]{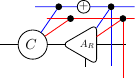}		
\end{gathered}
\end{equation}
together with
\begin{equation}
\label{tt_modified_center_definition}
\begin{gathered}
\includegraphics[valign=c]{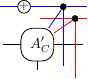}
=
\includegraphics[valign=c]{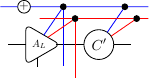}
=
\includegraphics[valign=c]{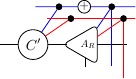}		
\end{gathered} .
\end{equation}
As Eq.~\eqref{tt_center_definition} is the gauge condition for the generalized MPS which is unique up to a phase we find that $A_C'\propto A_C$ and $C'\propto C$ which are the generalized VUMPS equations

\begin{equation}
\begin{gathered}
\frac{1}{\mathcal{N}}
\includegraphics[valign=c]{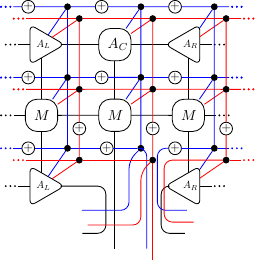}	
=	\tau_{A_C}
\includegraphics[valign=c]{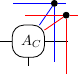}			
\end{gathered}
\end{equation}
and

\begin{equation}
\begin{gathered}
\frac{1}{\mathcal{N}'}
\includegraphics[valign=c]{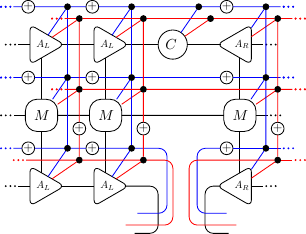}	
= \tau_C	
\includegraphics[valign=c]{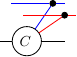}			
\end{gathered}
\end{equation}
which corresponds to Eqs.~\eqref{local_AC_eigenvalue_equation} and \eqref{local_C_eigenvalue_equation} and can be written in the normalized formulation as in Eq.~\eqref{local_h_a_c} and Eq.~\eqref{local_h_c}.
\end{widetext}

\section{Sequential implementation}\label{appendix:sequential_update}

The algorithm defined in the main text updates all tensors in parallel. Similarly, one could as well define a sequential update method. In particular, instead of updating the whole unit cell at once, one could iterate through the $x$-channel in Eqs.~\eqref{local_h_a_c} and~\eqref{local_h_c} as follows:

\begin{algorithm*}
\SetAlgoVlined
\SetInd{1em}{1em} 
\DontPrintSemicolon

\SetKwProg{Fn}{Function}{:}{}

\SetKwFunction{APAC}{local\_apply\_h\_a\_c}
\Fn{\APAC{$(A_C^{y})_y$, $x$}}{
$(A_C^{y})_y \longleftarrow $ update according to (\ref{local_h_a_c}) at $x$\;
\KwRet $(A_C^{y})_y$\;
}

\;
\SetKwFunction{APC}{local\_apply\_h\_c}
\Fn{\APC{$(C^{y})_y$, $x$}}{
$(C^{y})_y \longleftarrow $ update according to (\ref{local_h_c}) at $x$\;
\KwRet $(C^{y})_y$\;
}

\caption{Explicit terms of the local Hamiltonians in the sequential implementation.}
\label{environment_algorithms}
\end{algorithm*}

To start we initialize a set of $n_y$ MPS with a unit cell of size $n_x$, each of these MPS corresponding to a boundary. Then, for each $x$ in the MPS' unit cell we derive the local equations $h_{A_C}$ (cf.~ (\ref{local_h_a_c})) and $h_c$ (cf.~(\ref{local_h_c})) using (\ref{l_channel_fixed_Point}) and (\ref{r_channel_fixed_Point}) at $x$ (where $h_C$ actually should be derived for $x$ and $x-1$). Solving $h_{A_C}$ and $h_C$ we obtain new tensors $A_C$ at $x$ for all $y$ and $C$ at $x$ and $x-1$ for all $y$. Using these we update the boundaries using (\ref{nt_truncation_a_l}) and (\ref{nt_truncation_a_r}) obtaining a gauging error. Starting over again from the new set of boundaries for $x+1$ until one sweep through the unit cell is completed. Start again from the beginning until the gauging error (or the difference in the singular values of $C$ in different iterations) is smaller than a desired threshold $\epsilon$. The algorithm is shown with more structure in Algorithms \ref{environment_algorithms} and \ref{pvumps_algorithm}. 

The difference in the sequential and parallel update for local Hamiltonian quantum systems, similar to this work with $n_y=1$ and a hermitian MPO, is discussed in Ref.~\cite{ZaunerStauber2018} (cf. Figure 1). They find the two algorithms to perform comparably. While the parallel update has a smaller overhead per iteration computationally, because the environment terms need to be computed only once, the sequential algorithm seems to find short cuts in the iteration trajectory resulting in less iterations necessary till convergence. This can be understood nicely using the picture of Appendix \ref{appendix:non_trivial}: While the parallel update corresponds to the  minimization with respect to the full tangent space projector Eq.~\eqref{product_tangent_space_projector}, the sequential update  breaks these updates into optimizations with respect to hyperplanes within the tangent space (as the $x$-channel in the projector is fixed) with the consecutive updates in the $x$-direction being coupled. This means that for the sequential update there is more freedom to find different iteration trajectories, which however in general are not guaranteed to be shortevr. It is easy to see that a solution of one algorithm is also a solution to the other algorithm.

While the understanding carries over, the findings cannot be applied straightforwardly to this work. Here we are dealing with MPOs rather than local Hamiltonians. In turn, the environment equations are as complex as the leading terms in complexity, the computation of $A_C$. Therefore, the environment terms are a leading term in complexity such that the sequential implementation scales quadratically in $n_x$. Note: The sequential algorithm in Ref.~\cite{ZaunerStauber2018} scales quadratically in $n_x$, too, but only in a sub-leading term. For $n_x$ sufficiently larger than the physical dimension, however, the parallel update can be expected to perform better than the sequential update in their case. Obviously in our case it still holds that both algorithms converge to the same solutions.

\begin{algorithm*}
\SetAlgoVlined
\DontPrintSemicolon
\SetInd{1em}{1em} 

\KwData{MPO $M=(M_{x,y})_{x,y}$ with non-trivial $(n_x, n_y)$ shaped unit cell; desired accuracy $\epsilon_{prec}$}
\KwResult{Array containing the data of $n_y$ MPS corresponding to the boundaries $(\ket{\psi(A_y)})_y$}
\;
$\left(\ket{\psi(A_y)}\right)_y$ $\longleftarrow$ initialize MPS array\;
$(L_{x,y})_{x,y},\;\;(R_{x,y})_{x,y}\longleftarrow$ update environments from $\{\left(\ket{\psi(A_y)}\right)_y, M\}$  calling \ENvs\;
$\epsilon_{gauge} \longleftarrow 1$\;
$(\epsilon_{gauge, y})_y \longleftarrow (1)_y$\;
\While{$\epsilon_{gauge} > \epsilon_{prec}$}{
\For{$x\in \{0,\dots n_x\}$}{
$\left(A_C^{x,y}\right)_y \longleftarrow$ solve $h_{A_C}$ using an iterative solver calling \APAC at $x$\;

$\left(C_R^{x,y}\right)_y \longleftarrow$ solve $h_{C_R}$ using an iterative solver calling \APC at $x$\;

$\left(C_L^{x,y}\right)_y \longleftarrow$ solve $h_{C_L}$ using an iterative solver calling \APC at $x-1$\;	

\For{$y \in \{0,\dots, n_y\}$}{
$\ket{\psi(A_y)}, \epsilon_{trunc, y} \longleftarrow \{A_C^{x,y}, C_R^{x,y}, C_L^{x,y}\}$ following 
(\ref{nt_truncation_a_l}) and (\ref{nt_truncation_a_r})\;	
}		
$\epsilon_{gauge} \longleftarrow \text{max}\{\epsilon_{gauge, y}\}$\;
$(L_{x,y})_{x,y},\;\;(R_{x,y})_{x,y}\longleftarrow$ update environments from $\{\left(\ket{\psi(A_y)}\right)_y, M\}$  calling \ENvs\;
}
}
\caption{Sequential implementation of the VUMPS algorithm for non-trivial unit cells.}
\label{pvumps_algorithm}
\end{algorithm*}

\end{document}